\documentclass{aa}  

\PassOptionsToPackage{%
backend=bibtex8,bibencoding=ascii,%
language=auto,%
style=authoryear-comp,%
bibstyle=phys, 
maxbibnames=999,
maxcitenames=2,uniquelist=false, uniquename=false,
natbib=true 
}{biblatex}
\usepackage{xcolor}
\usepackage{graphicx}
\usepackage{scalerel}
\usepackage{comment}
\usepackage{multirow}
\usepackage{graphicx}
\usepackage{txfonts}
\usepackage{orcidlink}

\hypersetup{
    colorlinks=true,
    linkcolor=blue,
    citecolor=blue
    }
\newcommand{\degrees}{\ensuremath{^\circ}}

\begin{document}

   \title{A unique window into the Epoch of Reionisation: A double-peaked Lyman-$\alpha$ emitter in the proximity zone of a quasar at $z\sim 6.6$}
   \author{Klaudia Protušová\inst{1}\orcidlink{0009-0008-2205-7725},
   Sarah E.~I.~Bosman\inst{1,2}\orcidlink{0000-0001-8582-7012}, 
   Feige Wang\inst{3}\orcidlink{0000-0002-7633-431X},   
   Romain A.~Meyer\inst{4}\orcidlink{0000-0001-5492-4522},
   Jaclyn B.~Champagne\inst{5}\orcidlink{0000-0002-6184-9097},
   Frederick B.~ Davies\inst{2}\orcidlink{0000-0003-0766-2499},
   Anna-Christina Eilers\inst{6,7}\orcidlink{0000-0003-2895-6218},
   Xiaohui Fan\inst{5}\orcidlink{0000-0003-3310-0131},
   Joseph F.~Hennawi\inst{8,9}\orcidlink{0000-0002-7054-4332},
   Xiangyu Jin\inst{5}\orcidlink{0000-0002-5768-738X},   
   Hyunsung D.~Jun\inst{10}\orcidlink{0000-0003-1470-5901},
   Koki Kakiichi\inst{11}\orcidlink{0000-0001-6874-1321},
   Zihao Li\inst{11}\orcidlink{0000-0001-5951-459X},
   Weizhe Liu\inst{5}\orcidlink{0000-0003-3762-7344},
   Jinyi Yang\inst{3}\orcidlink{0000-0001-5287-4242}
   }

   \institute{
    Institute for Theoretical Physics, Heidelberg University, Philosophenweg 12, D–69120, Heidelberg, Germany
    \and
    Max-Planck-Institut für Astronomie, Königstuhl 17, 69117 Heidelberg, Germany 
    \and
    Department of Astronomy, University of Michigan, 1085 S.~University, Ann Arbor, MI 48109, USA
    \and
    D\'{e}partement d’Astronomie, University of Geneva, Chemin Pegasi 51, 1290 Versoix, Switzerland
    \and
    Steward Observatory, University of Arizona, 933 North Cherry Avenue, Rm. N204 Tucson, AZ 85721-0065, USA
    \and
    Department of Physics, Massachusetts Institute of Technology, Cambridge, MA 02139, USA
    \and
    MIT Kavli Institute for Astrophysics and Space Research, Massachusetts Institute of Technology, Cambridge, MA 02139, USA
    \and
    Department of Physics, University of California, Santa Barbara, CA 93106-9530, USA
    \and
    Leiden Observatory, Leiden University, P.O. Box 9513, 2300 RA Leiden, The Netherlands
    \and
    Department of Physics, Northwestern College, 101 7th Street SW, Orange City, IA 51041, USA 
    \and
    Cosmic Dawn Center (DAWN), Niels Bohr Institute, University of Copenhagen, Jagtvej 128, DK-2200 Copenhagen N, Denmark
    }
   \date{Received XXX; accepted YYY}

 
  \abstract
   {We present a detailed study of a double-peaked Ly$\alpha$ emitter, named LAE-11, found in the proximity zone of quasar J0910-0414 at $z \sim 6.6$. We use a combination of deep photometric data from Subaru Telescope, \textit{Hubble Space Telescope}, and \textit{James Webb Space Telescope} with spectroscopic data from Keck/DEIMOS, \textit{JWST}/NIRCam WFSS and \textit{JWST}/NIRSpec MSA to characterise the ionising and general properties of the galaxy, as well as the quasar environment surrounding it. Apart from Ly$\alpha$, we also detect H$\beta$, [O{\scriptsize III}]$_{\lambda\lambda4960,5008}$ doublet, and H$\alpha$ emission lines in the various spectral datasets. The presence of a double-peaked Ly$\alpha$ profile in the galaxy spectrum allows us to characterise the opening angle and lifetime of the central quasar as $\theta_Q>49.62\degrees$ and $t_Q > 3.8\times10^5$ years, and probe the effect of the quasar's environment on the star formation of the galaxy. LAE-11 is a fairly bright (M$_\mathrm{UV} = -19.84^{+0.14}_{-0.16}$), blue galaxy with a UV slope of $\beta = -2.61^{+0.06}_{-0.08}$ with moderate ongoing star formation rate (SFR$_\mathrm{UV} = 5.25\pm0.69~M_\odot$~yr$^{-1}$ and SFR$_\mathrm{H\alpha} = 12.93\pm1.20~M_\odot$~yr$^{-1}$). Since the galaxy is located in a quasar-ionised region, we have a unique opportunity to measure the escape fraction of Lyman Continuum photons using the un-attenuated double-peaked Ly$\alpha$ emission profile and its equivalent width at such high redshift. Moreover, we employ diagnostics which do not rely on the detection of Ly$\alpha$ for comparison, and find that all tracers of ionising photon leakage agree within 1$\sigma$ uncertainty. We measure a moderate escape of Lyman Continuum photons from LAE-11 of $f_\mathrm{esc}^\mathrm{LyC} = (8 - 33)\%$. Detections of both H$\alpha$ and H$\beta$ emission lines allow for separate measurements of the ionising photon production efficiency, resulting with $\log(\xi_\mathrm{ion}/\mathrm{Hz~erg^{-1}}) = 25.59\pm0.08$ and $25.65\pm0.09$, for H$\alpha$ and H$\beta$, respectively. The total ionising output of LAE-11, $\log(f_\mathrm{esc}^\mathrm{LyC}\xi_\mathrm{ion, H\alpha}/\mathrm{Hz~erg^{-1}}) = 24.85^{+0.24}_{-0.34}$, is higher than the value of $24.3 - 24.8$ which is traditionally assumed as needed to drive Reionisation forward. 
   }

   \titlerunning{A double-peaked Lyman-$\alpha$ emitter in the proximity zone of a quasar at $z\sim 6.6$}
   \authorrunning{K.~Protušová et al.}
   \maketitle
%

\section{Introduction} 

The Epoch of Reionisation (EoR) marks the second and last major phase transition in the history of the Universe. During this era, the majority of neutral hydrogen gas in the intergalactic medium (IGM) became ionised. Observations of the Lyman-$\alpha$ (Ly$\alpha$) forest in the spectra of high redshift quasars (QSOs) indicate that the EoR concludes at $z\sim 5.3$ \citep{Bosman_18,Kulkarni_19,Bosman_22, Gaikwad_23, Spina24}. Though the end of EoR is tightly constrained, its onset and course are still uncertain due to the unknown nature and output of ionising radiation (Lyman Continuum; $E > 13.6$ eV) of the sources driving reionisation of the IGM. 
The current leading candidates for major drivers of the EoR are early star forming galaxies, whose young and massive stars produce a large amount of Lyman Continuum photons (LyC, $\lambda < 912$\AA) \citep[e.g.][]{Robertson_13,Robertson_15,Finkelstein_19,Dayal_20,Naidu_22}.
\begin{figure*}[t]
  \includegraphics[width=\linewidth]{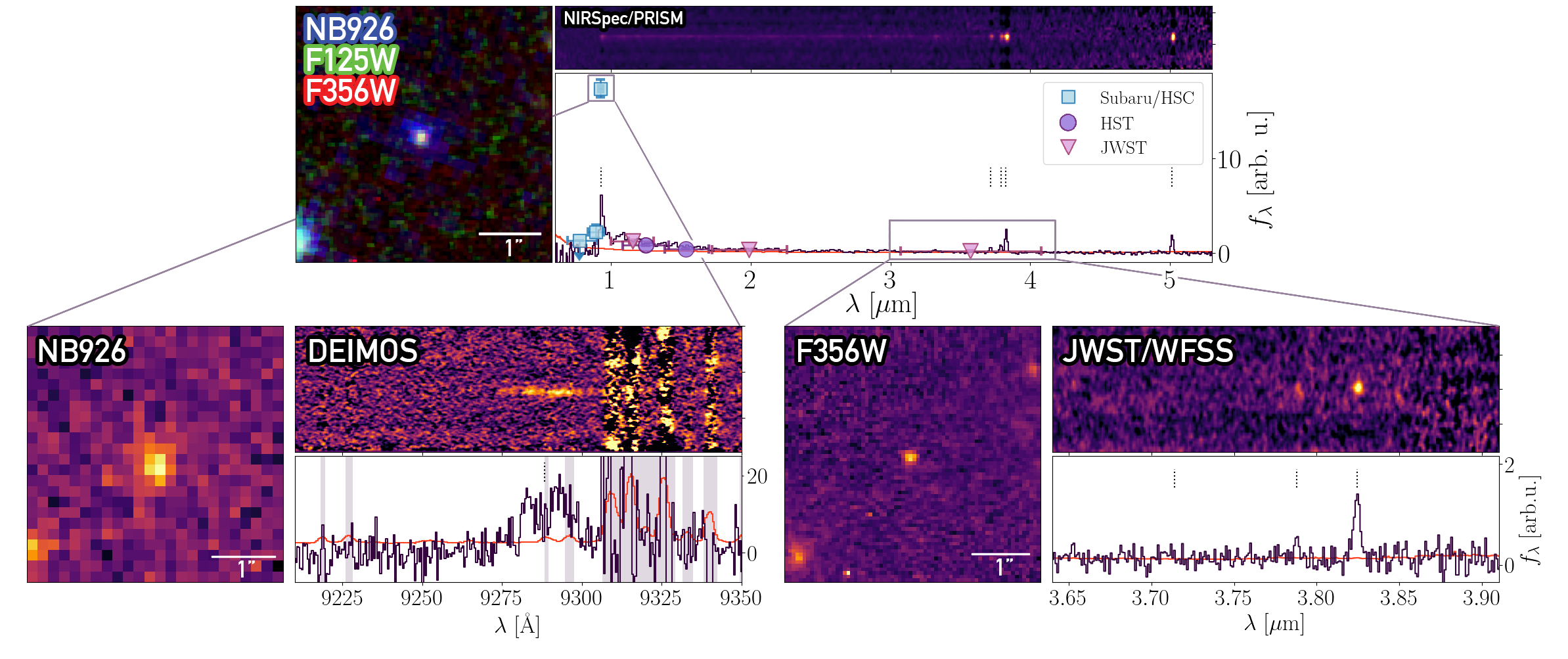}
  \caption{The available photometric and spectroscopic dataset of LAE-11. \textit{(upper)} A composite image of LAE-11 consisting of the \textit{NB926, F125W} and \textit{F356W} photometry, the 2D and 1D NIRSpec/PRISM spectrum of LAE-11 located at $z = 6.6405\pm0.0005$. The uncertainty on each pixel of the 1D spectrum is represented with a red line. The dotted vertical lines represent detected emission lines. From left to right: Ly$\alpha$, H$\beta$, [O{\tiny III}] doublet, and H$\alpha$. The spectral energy distribution of LAE-11, encompassing photometric data from Subaru/HSC (blue squares), \textit{HST}/WFC3 (purple circles), and \textit{JWST}/NIRCam (pink triangles) is shown on top of the 1D NIRSpec spectrum. \textit{(lower)} \textit{NB926} image, 2D and 1D Keck/DEIMOS spectrum of LAE-11 on the left with \textit{F356W} image, 2D and 1D \textit{JWST}/NIRCam WFSS spectrum of LAE-11 on the right. The uncertainty on each pixel in both spectra are represented with a red line. The dashed vertical lines mark the position of the Ly$\alpha$ in the DEIMOS spectrum, H$\beta$ and [O{\tiny III}] doublet emission lines in the NIRCam spectrum. The shaded vertical regions in the DEIMOS spectrum represent the skyline regions.
  }
  \label{fig:allDat}
\end{figure*}

With increasing redshift, the fraction of neutral hydrogen found in the IGM rises \citep[e.g.][]{Dijkstra_11, Inoue_14, Mason_18}, making direct observations of LyC escape from early star forming galaxies nearly impossible. For this reason, the nature of the galaxies contributing the majority of the ionising budget is still uncertain. Some research points towards rare luminous galaxies as major sources of ionising radiation \citep{Naidu_20, Nakane_24} as their late formation couples with the late onset of Reionisation. Others \citep[e.g.][]{Finkelstein_19, Mascia_24, Atek_24, Simmonds_24}, on the other hand, prefer the contribution from the numerous faint (M$_\mathrm{UV} > -18$) galaxies found at these times. Without measuring the amount of ionising photons these galaxy populations inject into the IGM, the character of the true drivers behind Reionisation cannot be easily resolved. 
 
While direct measurements of the LyC escape cannot be performed for high-$z$ sources, indirect diagnostics were inferred using low redshift ($z < 3$) analogues \citep[see][etc.]{Zackrisson_13, Zackrisson_17, Nakajima_14, Izotov_18, Jaskot_19, Katz_22, Flury_22a, Flury_22b, Choustikov_23}. Some of these indirect tracers link LyC escape with Ly$\alpha$ emission due to the high sensitivity of its radiative transfer to neutral hydrogen column density. Whenever Ly$\alpha$ photons encounter dense neutral gas, they scatter and Doppler-shift until they emerge out of resonance through the red or blue wing of the line shape, creating a double-peaked line profile. The separation of the two peaks is directly linked to the neutral hydrogen column density that blocks LyC photons from escaping along the line of sight to the observer \citep[e.g.][]{Verhamme_15, Kakiichi_Gronke_21}, and is therefore a powerful indirect tracer of LyC leakage \citep[e.g][]{Verhamme_15, Izotov_18, Flury_22b}. 


The double-peaked Ly$\alpha$ emission line profile is frequently observed in low-$z$ Ly$\alpha$ emitters (LAEs), such as Green Peas galaxies \citep[GPs;][]{Ortilova_18} and Ly$\alpha$ blobs \citep[LABs;][]{Matsuda_06}. However, the increased attenuation by the neutral IGM makes the detection of the blue peak rare at $z > 5$ \citep{Matthee_17, Shibuya_18, Mason_20, Tang_24b}, as even a small fraction of neutral gas ($x_{\mathrm{HI}}\geq10^{-4}$) leads to its suppression \citep{Dijkstra_14}. Despite their apparent rarity, a small number of high-$z$ LAEs residing in localised bubbles of ionised gas have been found, providing an excellent opportunity to measure the $f_{\mathrm{esc}}^{\mathrm{LyC}}$ of high-$z$ galaxies. These galaxies reside either in large ionised bubbles \citep{Hu_16, Matthee_18, Songaila_18, Meyer_21} or the proximity zones of high-redshift quasars \citep{Bosman_20}.


Quasar proximity zones trace small regions of highly ionised hydrogen around a luminous quasar, driven by the central AGN \citep{Madau_00, Wyithe_04, Eilers_17, Satyavolu_23}, allowing observations of flux blueward of Ly$\alpha$ at high-$z$. As seen in \citet{Zheng_06, Meyer_22, Wang_23} and others, high-$z$ quasars, due to their implied large halo masses, are expected to trace overdense regions and protoclusters of star-forming galaxies, potentially increasing the chance of detecting LAEs in their vicinity, however observational results do not reach a consensus with some quasars lacking an overdense environment \citep[e.g.][]{Banados_13, Mazzucchelli_17, Champagne_23, Rojas-Ruiz_24}. As such, a high number of LAEs have been found in the proximity zone of the quasar J0910-0414, which was part of the \textit{A SPectroscopic survey of biased halos In the Reionization Era} (ASPIRE) survey \citep{Wang_23}. One of these galaxies is located at a proper distance of $d_\mathrm{QSO}\sim 0.3$ pMpc from the quasar, designated LAE-11, and displays a double-peaked Ly$\alpha$ emission line \citep{Wang_24}, whose presence confirms the existence of an ionised region around J0910-0414. In this paper, we present a detail study of LAE-11 using data from Subaru Telescope, Keck Telescope, \textit{Hubble Space Telescope} (\textit{HST}), and \textit{James Webb Space Telescope} (\textit{JWST}). 

\setlength{\tabcolsep}{5.5pt}
\renewcommand{\arraystretch}{1.65}
\begin{table*}[t]
\caption{Photometry of LAE-11 obtained with the Subaru/HSC, \textit{HST}, and \textit{JWST}.}
\begin{tabular}{lcccccccc}
\hline
     & $i2$ & $z$ & $NB926$ & $F115W$ & $F125W$ & $F160W$ & $F200W$ & $F356W$ \\
    \hline\hline
    AB mag & $>27.83$ & $26.98\pm 0.34$ &$24.66\pm 0.06$ & $27.06\pm 0.15$ & $27.32\pm0.09$ & $27.53\pm 0.20$ & $27.17\pm 0.16$ & $26.39\pm 0.04$ \\
    $t_{\mathrm{exp}}$ [s] & $9000.00$ & $14400.00$ & $18780.00$ & $9582.75$ & $1417.25$ & $10897.40$ & $3779.34$ & $1417.25$\\
    \hline\end{tabular}
\label{tab:phot}
\end{table*}


The structure of this paper is as follows: In Section \ref{sec:obs} we describe the available dataset for LAE-11, and the reduction of each set of data. In Section \ref{sec:method}, we detail the procedure used to calculate the systemic redshift of LAE-11, methods used to extract emission line measurements and the UV properties of the galaxy, and the SED-fitting of the spectro/photometric data available in this study. In Section \ref{sec:results}, we present the measurements of the ionising properties of LAE-11, as well as the star formation rate (SFR) measurements. Moreover, we focus on the quasar itself and calculate its opening angle and lifetime. Finally, in Section \ref{sec:discussion}, we discuss the effects of the quasar ionising radiation on the galaxy, and discuss the impact that galaxies similar to LAE-11 had in the EoR. Throughout this work, we use the $\Lambda$CDM model \texttt{Planck18} as presented in \citet{Planck_20} with the following parameters: $\Omega_{\Lambda} = 0.69$, $\Omega_{\mathrm{m}} = 0.31$, and $H_0 = 67.7~\mathrm{km~s^{-1}Mpc^{-1}}$. All magnitudes quoted in this work are in the absolute bolometric (AB) system \citep{Oke_83}.

\section{Data}\label{sec:obs}
The quasar J0910-0414 was discovered by \citet{Wang_19} using the data from the DESI Legacy imaging Surveys \citep{desi}, the Pan-STARRS1 Survey \citep{panstarr} and additional infrared imaging data. Candidate selection was followed by spectroscopic observations, where J0910-0414 was observed by Magellan/FIRE \citep{magellan} and Gemini/GMOS \citep{gemini}. These spectra showcase a strong blueshifted C{\small IV} absorption feature, which categorises J0910-0414 as a Broad Absorption Line (BAL) quasar. The redshift of the host galaxy was measured with \textit{ALMA} to be $z = 6.6363 \pm 0.0003$ using the [C{\small II}] line \citep{Wang_inprep}. Due to its high supermassive black hole (SMBH) mass of $M_{\bullet} = (3.59\pm0.61)\times10^9~M_{\odot}$ measured from its Mg{\small II} line and UV brightness of M$_\mathrm{UV} = -26.61$ \citep{Yang_21}, this quasar was selected as a likely candidate to trace an overdensity of galaxies. The quasar field and consequently LAE-11 were observed with a number of photometric and spectroscopic surveys which we use in this work, described in the following Section. The entire available dataset for LAE-11 is summarised in Fig. \ref{fig:allDat}.

\subsection{Photometry}
\subsubsection{Subaru/HSC}
The first targeted photometric observations of the field surrounding quasar J0910-0414 were conducted from November 2019 to January 2020 with the Hyper Suprime-Cam (HSC) mounted on the 8.2 m Subaru Telescope \citep{hsc1,hsc2,hsc3,hsc4}. The imaging covered a circular field of view with $r\sim40\arcmin$ centred on the quasar \citep{Wang_24}. The field was observed using two broad band filters, $i2$ and $z$, and the narrow band filter $NB926$ designed to capture the Ly$\alpha$ emission of J0910-0414's surrounding galaxies across the redshift range $6.494 < z_{\mathrm{Ly\alpha}} < 6.739$. The exposure times were $\sim$150 minutes, $\sim$240 minutes, and $\sim$313 minutes, respectively. 

The detailed data reduction process using the \texttt{hscPipe 6.7} pipeline \citep{Juric_17,Bosch_18} and LAE selection criteria are described in \citet{Wang_24}, though we provide a brief overview here. First, each individual observation was processed, calibrated, and then combined into a stacked mosaic. Afterwards, the pipeline implements photometric and astrometric calibrations to the stacked mosaics, resulting in a seeing in each mosaic of 0.79\arcsec, 0.74\arcsec, and 0.56\arcsec~with $5\sigma$ limiting magnitudes measured in 1\arcsec.5 diameter apertures of 27.31, 26.27, and 25.71 for filters $i2,~z,\mathrm{~and~}NB926$, respectively. To create a dataset of nearby LAE candidates, \citet{Wang_24} imposed a colour selection requiring a substantially stronger detection in the narrow band filter compared to the broad band filter. The photometry was then extracted using the \texttt{hscPipe} pipeline. We use the measured magnitudes found in \citet{Wang_24}



\subsubsection{\textit{HST}}
To measure the rest-frame UV emission and spectral energy distribution (SED) of the galaxies clustered around the quasar J0910-0414, we use photometry with  high spatial resolution which was obtained using the Wide Field Camera 3 \citep[WFC3;][]{wfc3} on board of the \textit{HST} \citep[][PID 16187; P.I. Feige Wang]{HST_prop} on March 10th, 2022. The observations utilised 4 primary orbits and were taken in the NIR \textit{F125W} and \textit{F160W} bands, centred on the position of the quasar, covering an area of $\sim31~\mathrm{arcmin}^2$. The \textit{HST} imaging was processed, calibrated and drizzled using the \texttt{DrizzlePac} package \texttt{v3.2.1} \citep{HST_pipeline} with the reference calibration file \texttt{hst\_0988.pmap}. The resulting drizzled and stacked mosaics result in exposure times of 9582.75~s and 10897.40~s with 5$\sigma$ limiting magnitudes of 28.26 and 27.98 measured in a 0\arcsec.4 diameter apertures for filters \textit{F125W} and \textit{F160W}, respectively. We extracted the photometry using \texttt{source-extractor} \citep{Bertin_96}.

\subsubsection{\textit{JWST}}
Additionally, we use complementary data obtained with the NIRCam instrument \citep{nircam} on board of the \textit{JWST}, which were taken as part of the medium-size \textit{JWST} Cycle 1 program ASPIRE (PID 2078, P.I. Feige Wang). The survey focused on 25 quasar fields at $6.5< z <6.8$, including the quasar J0910-0414. The photometric observations were taken in August 2022 with the broad band filters \textit{F115W}, \textit{F200W}, and \textit{F356W}, covering an area of $\sim11~\mathrm{arcmin}^2$. This configuration covers the rest-frame optical emission from the quasar and its surrounding galaxies. The NIRCam data were processed using the standard \textit{JWST} pipeline \citep{Bushouse_22} \texttt{v1.8.3} in combination with a custom set of scripts which are detailed in \citet{Wang_23} and \citet{Yang_23}. The reference calibration file \texttt{jwst\_1015.pmap} was utilised. The resulting exposure times are 1417.254~s, 3779.344~s, and 1417.254~s and 5$\sigma$ magnitudes of 29.24, 29.68, and 27.82 measured in 0\arcsec.32 diameter apertures for filters \textit{F115W}, \textit{F200W}, and \textit{F356W}, respectively. A source catalogue was extracted using \texttt{SourceXtractor++} \citep{se++} and the detailed procedure is described in \citet{Champagne_24a}.

\subsection{Spectroscopy}
\subsubsection{Keck/DEIMOS}
We used optical spectroscopic data provided by the DEep Imaging Multi-Object Spectrograph (DEIMOS) \citep{deimos} mounted on the 10 m Keck-II Telescope, which were obtained as a spectroscopic follow-up of the LAE candidates in the field of J0910-0414 by \citet{Wang_24}. LAE-11 was observed on the 25 Feb 2022, utilising a mask with a $\sim 1\arcsec$ slit width, the filter \textit{OG550}, and the 830G grating \citep{Wang_24}. The combination of these parameters allows for a spectral dispersion of $0.47$~\AA/px and a resolving power of $R \sim 4650$. The seeing on the night of the observation varied from 0.6\arcsec to 1.5\arcsec. Overall, 18 exposures lasting 1200~s were taken. 

We reduced the available spectroscopic data using the Python-based package for semi-automated reduction of astronomical spectra \texttt{Pypeit v1.15.1} \citep{Prochaska_20}. We follow a standard spectroscopic reduction for Keck/DEIMOS, consisting of flat fielding, wavelength calibration, optimal extraction, and flux calibration for each observation. We carefully model the sky lines to correctly reconstruct the underlying spectrum. The normalised residuals of the sky model are presented in Fig. \ref{fig:skysub}. Afterwards, the 1D and 2D reduced spectral outputs were stacked, resulting in the total exposure time of 6 hours. 

\begin{figure}[t]
    \centering
    \includegraphics[width=\linewidth]{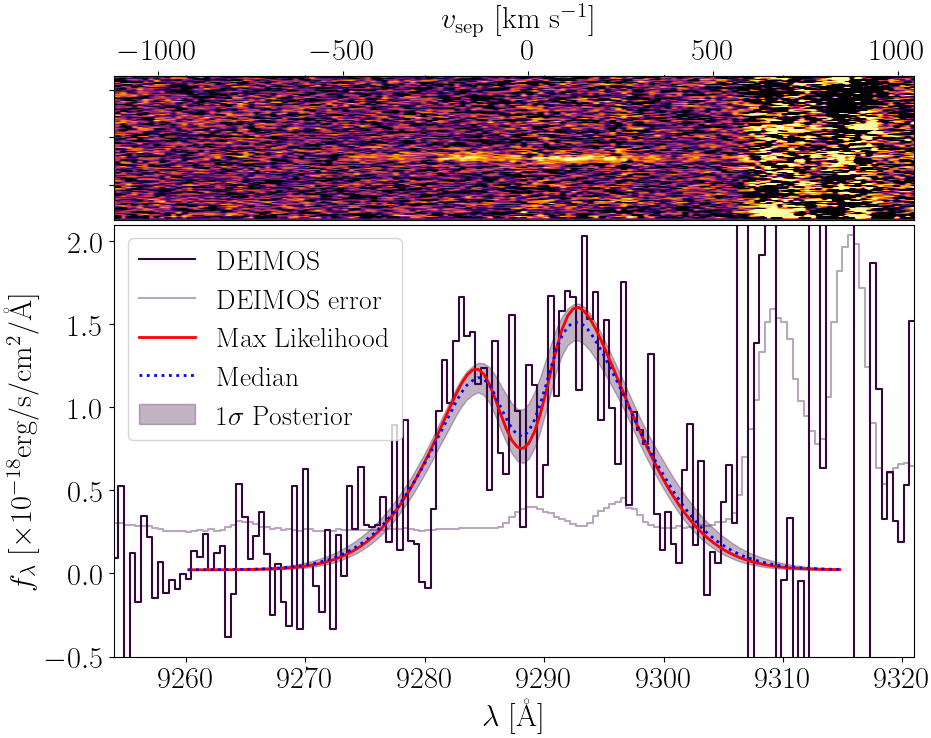}
    \caption{(\textit{upper}) 2D spectrum showcasing the double-peaked Ly$\alpha$ emission. (\textit{lower}) The MCMC fit to the Ly$\alpha$ line profile. The dark and light purple lines mark the observed spectrum and its uncertainty. The red line indicates the maximum likelihood model. The dotted blue line shows the median model and the shaded region marks its $1\sigma$ posterior.}
    \label{fig:LyaProf}
\end{figure}

\subsubsection{\textit{JWST}/NIRCam WFSS}
In addition to imaging survey, the ASPIRE collaboration obtained spectroscopic observations of the targeted 25 quasar fields using NIRCam \textit{Wide Field Slitless Spectroscopy} (WFSS). These observations were targetting the [O{\small III}] doublet of galaxies in the redshift range of $5.3 < z < 6.5$. The full description of the reduction steps is described in detail in \citet{Wang_23}, but we provide a brief overview here. The WFSS observations employed grism-R paired with the \textit{F356W} filter, providing a wavelength coverage spanning $3.1-3.9~\mu$m. The detection of LAE-11 fell onto module A of the detector. A primary dithering pattern was employed, using the three-point \texttt{INTRAMODULEX} pattern, with a secondary subpixel dithering pattern \texttt{2-POINT-LARGE-WITH-NIRISS}. The data were reduced using the same procedure as the NIRCam imaging data. Additionally, a tracing and dispersion model based on the work of \citet{Sun_22} was built to extract the 1D spectra from individual exposures with optimal extraction \citep{Horne_86}. These 1D spectra were then stacked with inverse variance weighting as detailed in \citet{Wang_23}. However, \citet{Wang_23} note that the offset between the spectral tracing model and the data along the dispersion axis cannot be estimated due to the lack of in-flight wavelength calibration data. As a conservative measure, they propose a constant velocity offset of $<100$ km~s$^{-1}$, or $\Delta z < 0.003$ for the detected [O{\small III}] emitters. The final spectra have exposure times of 3779.343~s, resolving power of $\sim$1300–1600 and dispersion of $\sim 10$~\AA/px. 



\subsubsection{\textit{JWST}/NIRSpec MSA}
Finally, we use data from a follow-up observation of J0910-0414 and its protocluster with \textit{JWST}/NIRSpec \textit{MultiObject Spectroscopy} (MSA) \citep{nirspec}, which was conducted in November 2023 (PID 2028; P.I. Feige Wang). The \textit{PRISM/CLEAR} setup was employed in order to target the rest-frame UV and optical part of the spectrum of galaxies at $z\sim6.6$ surrounding the quasar. The NIRSpec spectrum covers a wavelength range of $\lambda = (6000 - 53000)$ \AA, or the rest-frame wavelength range of $\lambda_\mathrm{rest} = (785 - 6940)$ \AA. The data was reduced and processed using the standard \textit{JWST} pipeline \texttt{v1.13.3} and calibration reference file \texttt{jwst\_1015.pmap} was used in the process. The exposure time of the final spectrum is 3939.0~s and has a resolution of $R\sim100$ with an average dispersion of $\sim100$ \AA/px.

\begin{figure}[t]
    \centering
    \includegraphics[width=\linewidth]{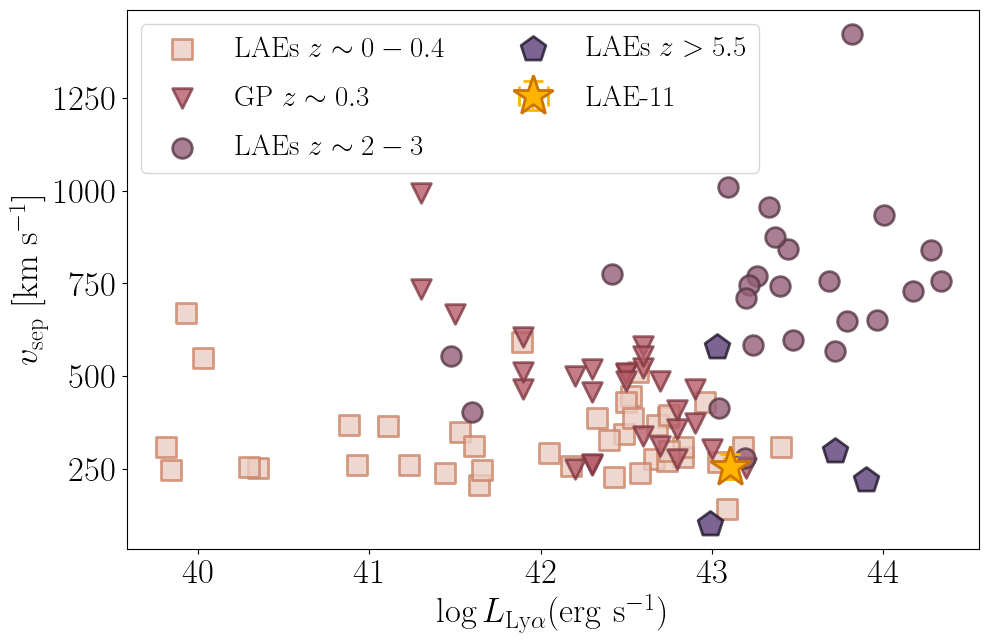}
    \caption{The velocity separation of the Ly$\alpha$ emission profile in relation to its luminosity for $z\sim0 - 0.4$ LAEs \citep{Izotov_18, Izotov_20, Izotov_21, Izotov_22, Izotov_24}, $z\sim0.3$ GPs \citep{Yang_17}, $z\sim2-3$ LAEs \citep{Kulas_12,Hashimoto_15, Vanzella_16}, and $z\geq5.5$ LAEs \citep{Hu_16, Songaila_18,Bosman_20,Meyer_21}. LAE-11 is most consistent with local luminous GPs and LAEs.}
    \label{fig:logLvsep}
\end{figure}

\section{Methodology}\label{sec:method}
In this section we first describe the methods used to constrain the redshift of LAE-11 and measure the properties of various spectral lines present in the available data. Next, we describe the model and process of SED fitting of the spectrophotometric data. Finally, we measure the UV slope $\beta_{1550}$ and absolute UV magnitude M$_\mathrm{UV}$ of the galaxy, as well as its dust properties.

\subsection{Redshift calibration}\label{sec:redshift}
We estimate the systemic redshift of LAE-11 by determining the position of the trough of the double-peaked Ly$\alpha$ profile and by determining the redshift of the detected [O{\small III}] doublet. We find a $1\sigma$ discrepancy between these two  redshift estimates. We attempt to mitigate this tension by turning to the other available LAEs which are also [O{\small III}] emitters in our sample of galaxies.
 
We start by calculating the systemic redshift of LAE-11 using the trough of the double-peaked Ly$\alpha$ profile in the DEIMOS spectra, as it has been found to be a reliable tracer of systemic redshift \citep[e.g.][]{Verhamme_06,Gronke_15,Verhamme_18}. We use an MCMC analysis Python package \texttt{emcee} \citep{emcee} to fit the Ly$\alpha$ profile with a combination of a simple emission Gaussian and an absorption Voigt profile parameterised by \citet{Garcia_06}. The free parameters describe the amplitude A$_\mathrm{Gauss}$, redshift $z_\mathrm{Ly\alpha}$ and variance $\sigma_\mathrm{Gauss}$ of the Gaussian component, and the redshift $z_\mathrm{trough}$, neutral hydrogen column density $N$ and the broadening parameter of the Voigt profile $b$. We initialise the fit with 500 Monte Carlo chains each undergoing 5000 iterations in order to sample the entire prior parameter space. To minimise the effect of the prior conditions on the best-fit parameters, we discard the first 1000 steps of each chain in the burn-in phase. The best-fit parameters and their errors are then calculated as the mean and 1$\sigma$ of the combined MCMC chain steps. The priors, initial and best-fit values for each parameter can be found in Table \ref{tab:LyA}, with the best-fit model presented in Fig. \ref{fig:LyaProf}. We measure the redshift of the trough to be $z = 6.6405\pm 0.0005$, in line with the measurement presented in \citet{Wang_24}. 

\begin{table}[t]
\caption{The parameters, their priors, and fitted values of the line profile model using a combination of simple emission Gaussian and an absorption Voigt profile, which was employed in the MCMC fitting of the Ly$\alpha$ emission line profile.}
\centering\label{tab:LyA}
    \begin{tabular}{lccc}
    \hline
    Parameter & Prior & Value\\ \hline\hline
    log A$_\mathrm{Gauss}$ [cgs] & $-17> \log \mathrm{A_{Gauss}} > -20$& $-17.76\pm0.07$ \\ 
    $\sigma_\mathrm{Gauss}$ [\AA] & > 10 & $6.89\pm0.74$ \\
    $z_\mathrm{Ly\alpha}$ & $6.63 < z_\mathrm{Ly\alpha} < 6.65$ & $6.6416\pm0.0004$ \\
    $z_\mathrm{trough}$ & $6.635 < z_\mathrm{trough} < 6.647$ & $6.6405\pm0.0005$  \\
    log $N$ [cm$^{-2}$] & $ 12< \log N< 15$ & $13.96\pm0.24$ \\
    $\log b$ [cm/s] & $6.5< \log b < 7.5$ & $6.93\pm0.12$ \\ \hline
    \end{tabular}
\end{table}

\begin{figure}
    \centering
    \includegraphics[width=\linewidth]{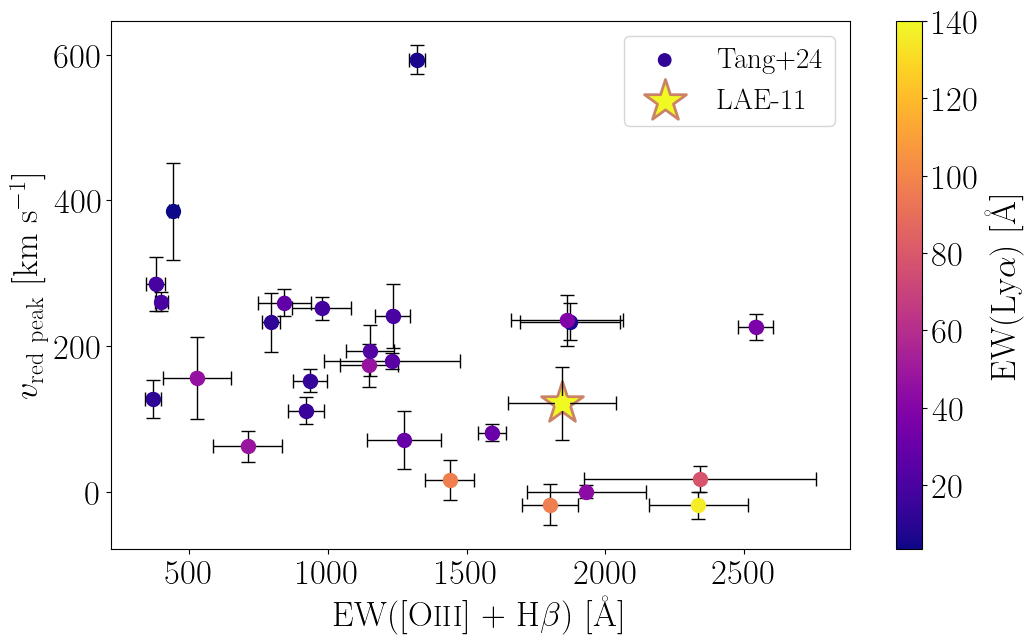}
    \caption{The correlation between [O{\tiny III}] + H$\beta$ EW$_0$ and the velocity offset of the red Ly$\alpha$ peak as measured by \citet{Tang_24} for 26 EELGs at $z\sim2 - 3$ (circles). LAE-11 is depicted with a star. The colour of each data point corresponds to the Ly$\alpha$ EW$_0$ of each galaxy.}
    \label{fig:o3hbvpeak}
\end{figure}

We confirm the redshift calculation following the method of \citet{Verhamme_18}. In this approach, the correlation between the velocity offset of the red peak $v_\mathrm{red}$ with respect to the systemic redshift, and half of the separation of the two peaks can be described with an almost one-to-one relation. We determined the peak separation by translating the fitted Ly$\alpha$ emission line model in velocity space and determining the local maxima. We find the velocity separation of the two peaks is $\Delta v_\mathrm{sep} = 255.71\pm33.69$~km s$^{-1}$. The velocity separation is comparable to local bright GPs and LAEs (see Fig. \ref{fig:logLvsep}). Using the relation from \citet{Verhamme_18}: 
\begin{equation}
    v_\mathrm{red} = (1.05\pm0.11)\times\Delta v_{\mathrm{sep},1/2} - (12\pm37)\mathrm{~km~s^{-1},}
\end{equation} we find $v_\mathrm{red} = 122.25 \pm 47.79$ km s$^{-1}$, or $z = 6.640 \pm 0.001$, which is in agreement with our initial measurement. The offset of the red peak is comparable to Extreme Emission Line Galaxies (EELGs) examined in \citet{Tang_24}, where galaxies with high equivalent widths of ([O{\small III}] + H$\beta$) and Ly$\alpha$ show small peak offset (see Fig. \ref{fig:o3hbvpeak}).

Next we turn to the JWST spectra found in our data sample, specifically to the NIRCam WFSS due to its high resolution. We identify a doublet in the 2D and 1D spectra which would correspond to the [O{\small III}] doublet at $z\sim 6.6$. We fit each line with a simple Gaussian and tie the peak separation as dictated by atomic physics \citep{o3ratio}. The [O{\small III}] doublet suggests that LAE-11 is located at $z = 6.6372 \pm 0.0004$. This redshift estimation is in tension with our Ly$\alpha$ trough calculation even when adopting the conservative calibration redshift offset $\Delta z < 0.003$ quoted by \citet{Wang_23}, resulting in 1$\sigma$ tension. This redshift discrepancy cannot be resolved using the NIRSpec data, as the wavelength resolution is too low to differentiate between the two redshift solutions.

\setlength{\tabcolsep}{5.5pt}
\renewcommand{\arraystretch}{1.65}
\begin{table*}[t]
\caption{The measured emission line properties in the spectra of LAE-11. EWs and their uncertainties are given in \AA. Fluxes and flux errors are measured in $\times10^{-18}$ erg/s/cm$^2$. Luminosities and luminosity uncertainties are given in $\times10^{41}$ erg/s.}
\centering
    \begin{tabular}{lccc|ccc}\hline
    \multicolumn{1}{c}{} & \multicolumn{3}{c|}{DEIMOS \& NIRCam/WFSS} & \multicolumn{3}{c}{NIRSpec/PRISM} \\
      Line & EW$_0$ & $F_\lambda$ & $L_\lambda$ & EW$_0$ & $F_\lambda$ & $L_\lambda$ \\ \hline\hline
      
      Ly$\alpha$ & $140\pm10$ & $25.34\pm1.57$ & $127.72\pm9.48$ & $41\pm11$& $3.72\pm1.45$ & $19.57\pm7.64$ \\
      H$\beta$ & --- & --- & --- & $157\pm53$ & $1.26\pm0.18$ & $6.62\pm0.96$ \\
      
      [O{\small III}]$_{4960}$ & $426\pm121$ & $1.75\pm0.24$ & $9.18\pm1.24$ & $462\pm73$ & $1.95\pm0.17$ & $10.26\pm0.89$\\
      
      [O{\small III}]$_{5008}$ & $1172\pm157$ & $5.07\pm0.51$ & $26.68\pm2.27$ & $1225\pm170$ & $5.57\pm0.21$ & $29.30\pm1.10$\\
      H$\alpha$ & --- & --- & --- & $1483\pm246$ & $3.11\pm0.29$ & $16.36\pm1.52$ \\ \hline     
    \end{tabular}\label{tab:lines}
\end{table*}

We attempt to correct this discrepancy by turning to the other LAEs reported by \citet{Wang_24} which are present in the ASPIRE data. These galaxies are LAE-1 ($z_\mathrm{[OIII]} = 6.537\pm0.003$) and LAE-12 ($z_\mathrm{[OIII]} = 6.632\pm0.003$). Both galaxies showcase a Ly$\alpha$ profile with only one peak. We use the relation found by \citet{Verhamme_18}, which ties the full width half maximum (FWHM) of the Ly$\alpha$ line and the offset of the red peak calibrated on a sample of low-$z$ LAEs which are also C{\small III}], H$\alpha$ or [O{\small III}] emitters:
\begin{equation}\label{eq:fwhm}
    v_\mathrm{red} = (0.9\pm0.14)\times\mathrm{FWHM(Ly\alpha)} - (34\pm60)~\mathrm{km~s^{-1}.}
\end{equation}
We fit the spectrum of LAE-1 with a spline due to the asymmetric shape of the Ly$\alpha$ line and a Gaussian to the Ly$\alpha$ line of LAE-12 to determine their FWHM. Using Eq. \ref{eq:fwhm}, we find that the velocity offset for LAE-1 and LAE-12 is $v_\mathrm{red} = 143.30 \pm 66.04$ km s$^{-1}$ and $v_\mathrm{red} = 57.03 \pm 61.65$ km s$^{-1}$, respectively. These offsets are then converted to redshifts as $z = 6.538\pm 0.002$ and $z = 6.640\pm 0.002$, for LAE-1 and LAE-12. We perform the same procedure using the right peak of LAE-11's Ly$\alpha$ profile and find $z = 6.640\pm 0.002$. For LAE-1, we find that this redshift agrees with the value determined using its [O{\small III}] doublet as $z_\mathrm{Ly\alpha} - z_\mathrm{[OIII]}$ = 0.001, however for LAE-12 the value is in 2.6$\sigma$ tension with its [O{\small III}] redshift with $z_\mathrm{Ly\alpha} - z_\mathrm{[OIII]}$ = 0.008. Meanwhile LAE-11 once again displays 1$\sigma$ tension of $z_\mathrm{Ly\alpha} - z_\mathrm{[OIII]}$ = 0.003. 

\citet{Torralba_24} showcase an offset in wavelength calibration present in spectra of the same object depending on the NIRCam module used for the observation. We check the modules of our galaxy sample and find that both galaxies with redshift discrepancies fall on module A of the NIRCam detector, while LAE-1 is located on module B (see Fig. \ref{fig:module}). This would imply calibration issues with module A, however \citet{Torralba_24} find the calibration issues on module B. For this reason, we adopt the redshift determined from the trough in the Ly$\alpha$ profile measured from its model as the systemic redshift. However, we discuss the implications on the ionising properties of LAE-11 when taking $z_\mathrm{[OIII]}$ at face-value in Sec. \ref{sec:o3redshift}.

\subsection{Emission line measurements}
Due to the high sensitivity of the NIRSpec detector, we are able to detect the continuum emission in the rest-frame UV, and marginally in the rest-frame optical region. To establish the continuum flux density in the UV region of the spectrum, we fit a simple power law from Ly$\alpha$ line to the Balmer break. Due to the <2$\sigma$ detection of the continuum from the Balmer break to the end of the spectrum, we use the continuum flux density calculated by the SED fitting (see Sec. \ref{sec:sed}). 

Afterwards, we model and fit each detected line using a simple Gaussian model and the power-law continuum determined above. We model the high-resolution spectrum of the Ly$\alpha$ line with a Gaussian multiplied by a Voigt profile as described previously. We also fit each peak with a simple Gaussian to find the red-to-blue peak flux ratio ($R/B$), which results in $R/B \sim 1.2$. When fitting the [O{\small III}] doublet, we fit two Gaussians simultaneously. We tie the peak separation of the two Gaussians, however we leave the redshift of the two peaks as a free parameter. Additionally we impose a uniform prior on the expected flux ratio of the lines from $2.48$ to $3.48$, centred on $2.98$ as dictated by atomic physics \citep{o3ratio}. The uncertainties in the measured properties are determined by perturbing the spectrum by a random draw from a Gaussian distribution whose standard deviation is set to the observed uncertainty of the spectrum. We repeat this process 1000 times, and calculate the 1$\sigma$ standard deviation for each property. The detected lines, their fluxes, luminosities and equivalent widths (EWs) can be found in Table~\ref{tab:lines}.

\begin{figure*}[t]
    \centering
    \includegraphics[width=\linewidth]{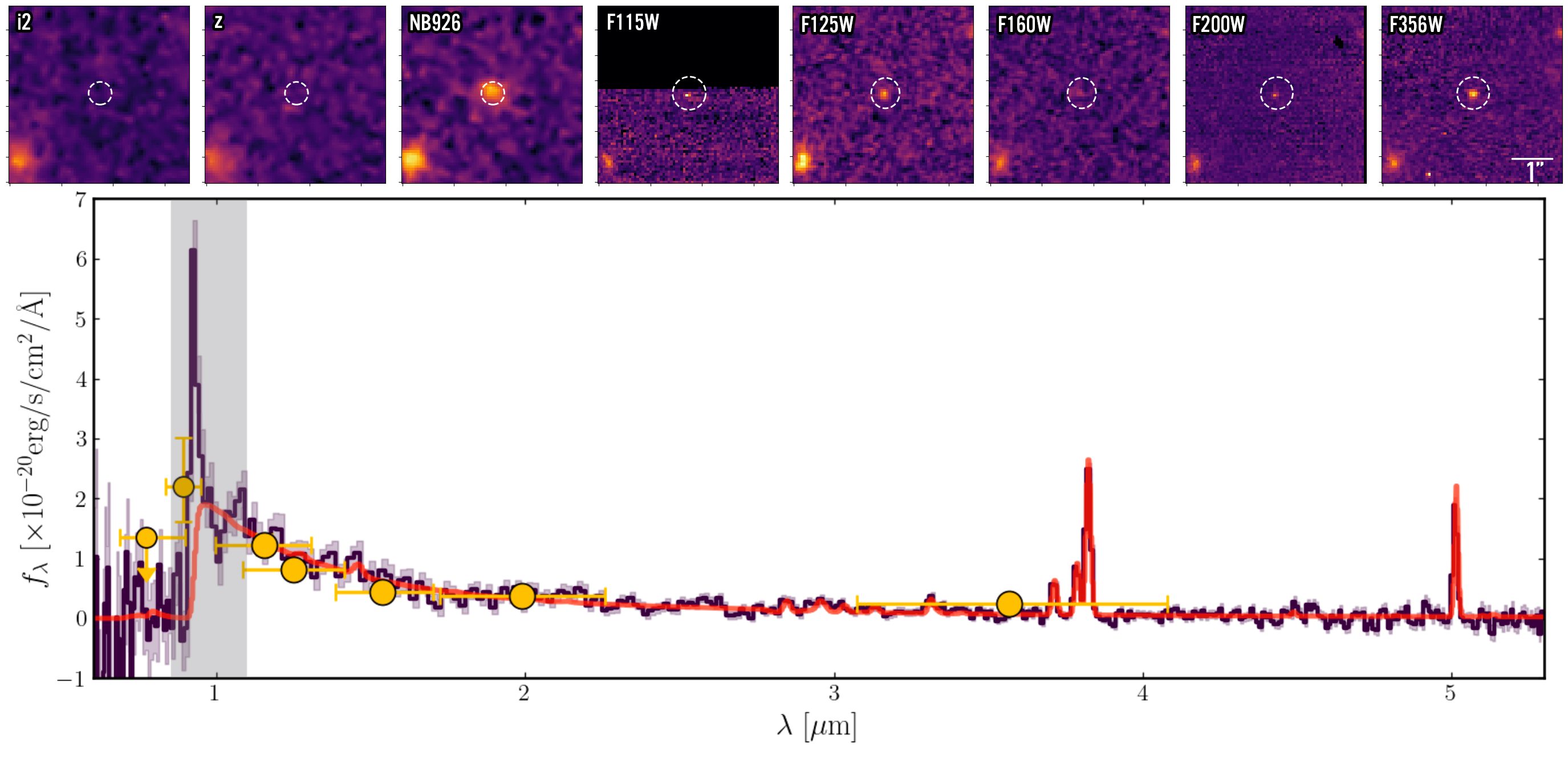}
    \caption{\textit{(upper)} Imaging of LAE-11 in all available bands across Subaru/HSC, \textit{HST}/WFC3 and \textit{JWST}/NIRCam, sorted by wavelength. The white dashed circle corresponds to the aperture used to measure the photometry in each filter. \textit{(lower)} The observed NIRSpec PRISM spectrum and its 1$\sigma$ uncertainty are depicted with a dark purple line and shaded region, the observed photometry with yellow points and the best-fitting SED with a red line. The shaded grey region marks the masked Ly$\alpha$ emission line, which was excluded from the fitting process due to the limitations of the model.}
    \label{fig:sed}
\end{figure*}

\subsection{UV slope $\beta_{1550}$, rest-frame UV magnitude M$_\mathrm{UV}$, and dust properties}
Since we have access to the UV continuum emission in the NIRSpec data, we are able to directly measure the UV slope $\beta_{1550}$ of LAE-11. The UV slope $\beta_{1550}$ is parametrised as a slope of a power-law, following the relation $f_\lambda \propto \lambda^\beta$ at $\lambda_\mathrm{rest} = 1550$~\AA~following \citet{Calzetti_94}. We measure the UV slope $\beta_{1550}$ by fitting a power-law to the observed spectrum in 10 wavelength windows, which are defined in \citet{Calzetti_94}. To estimate the uncertainty on the measured $\beta_{1550}$, we perturb the observed spectrum 1000 times as before. Each perturbed spectrum is then fitted within the same region with a power-law. The final uncertainty is then calculated as the standard deviation of the distribution of measured $\beta_{1550}$ of the 1000 values. The measured UV slope of the NIRSpec spectrum of LAE-11 is $\beta_{1550} = -2.61^{+0.06}_{-0.08}$. 


In addition to UV slope, we are able to calculate the absolute rest-frame UV magnitude. We use the measured photometric flux from filter \textit{F115W}, as it is centred on $\lambda_\mathrm{rest} = 1500$~\AA. We convert the photometric flux to absolute UV magnitude using the luminosity distance of the galaxy. Using this method, we find that the absolute UV magnitude is M$_\mathrm{UV,phot} = -19.84^{+0.14}_{-0.16}$. We also measure the absolute UV magnitude using the fit to continuum emission performed in previous section, which results in M$_\mathrm{UV,cont} = -19.90^{+0.13}_{-0.14}$. These two values agree within 1$\sigma$ uncertainty. Throughout the paper, we will use the M$_\mathrm{UV}$ calculated from photometry. The resulting values can be found in Table \ref{tab:properties}.

Next, we look at the dust properties of LAE-11. First, we use the tight correlation determined by \citet{Chisholm_22}:
\begin{equation}
    E(B-V) = (0.47\pm0.01) + (0.171\pm0.007)\times\beta_{1550}
\end{equation}
to convert the UV slope $\beta_{1550}$ to colour excess $E(B-V)_\mathrm{UV}$, resulting in $E(B-V)_\mathrm{UV} = 0.02_{-0.02}^{+0.03}$. Following the relation between colour excess and extinction described in \citet{Dominguez_13}:
\begin{equation}
    A_V = (4.05\pm0.80)\times E(B-V)\mathrm{~,}
\end{equation}
we find $A_V = 0.08_{-0.08}^{+0.14}$, suggesting that LAE-11 is a dust-poor system. This is further supported by the dust attenuation obtained from SED fitting ($A_V<0.01$) described in the following Section.


\subsection{SED fitting with \texttt{BAGPIPES}}\label{sec:sed}
To constrain the properties of the stellar population of the galaxy we perform SED fitting with the \texttt{BAGPIPES} code \citep{bagpipes}, using a combination of NIRSpec/PRISM spectroscopy and photometric data. \texttt{BAGPIPES} employs the stellar models introduced in \citet{Bruzal_03}, which utilise the empirical spectral library MILES \citep{Falcon-Barroso_11} and \citet{Kroupa_02} initial mass function (IMF). Additionally, the models are complemented with additional nebular and continuum emission based on \texttt{Cloudy} models \citep{Ferland_98, Byler_17}. 

The redshift of LAE-11 is fixed to the systemic redshift determined by the trough of the double-peaked Ly$\alpha$ profile to $z = 6.6405$. We adopt the non-parametric star formation history (SFH) with $N_\mathrm{SFH~bins} = 7$ bins of \citet{Leja_19}, with the first four spanning $0-2$~Myr, $2-4$~Myr, $4-8$~Myr, and $8-16$~Myr in lookback time to be sensitive to the variation in the recent SFH of the galaxy. The remaining bins are logarithmically spaced in the galaxy's lookback time from 16~Myr to $z = 20$ \citep[e.g.][]{Tacchella_22}. Finally, we apply dust attenuation following the dust model introduced in \citet{Calzetti_94} and allow $A_V$ to uniformly vary as $0<A_V <4$. For the remaining parameters (stellar mass $M_\star$, ionisation parameter $U$, metallicity $Z$) we use broad uniform priors, listed in Table \ref{tab:sed}. We mask the wavelength region spanning $1150$ \AA~$<\lambda_\mathrm{rest}< 1450$ \AA, due to the limitations of the model to reproduce the Ly$\alpha$ region. The resulting best-fit SED template, the predicted and observed photometry, and the NIRSpec spectrum are presented in Fig. \ref{fig:sed}. The best-fit SFH of LAE-11 is shown in Fig. \ref{fig:sfh}. The best-fit values of the physical parameters are shown in Table \ref{tab:properties}. The posterior distributions for each parameter can be found in Fig. \ref{fig:bagpipesPosterior}.

\section{Results}\label{sec:results}
In this Section, we evaluate the properties of LAE-11. First, we calculate the escape fraction of Lyman Continuum photons using a set of proxies based on empirical relations and multivariate-empirical models. Next, we focus on the ionising photon production efficiency of the galaxy. Afterwards, we measure the star formation rate of LAE-11. Finally, we investigate the quasar-galaxy geometry and the quasar environment of J0910-0414. 

\subsection{Escape fraction of Lyman Continuum $f_\mathrm{esc}^\mathrm{LyC}$}
To investigate the degree of LyC leakage undergoing in LAE-11, we must evaluate $f_\mathrm{esc}^\mathrm{LyC}$, the fraction of ionising photons which are able to escape into the IGM. Due to the low IGM attenuation of the Ly$\alpha$ line of high-$z$ galaxies situated in quasar proximity zones, we have a unique opportunity to use the EW and peak separation of Ly$\alpha$ profile to estimate LyC leakage of LAE-11. 


First, we measure the escape fraction of LyC photons using the peak separation of the Ly$\alpha$ profile \citep{Verhamme_17, Izotov_18, Flury_22b}. The empirical correlation between the velocity separation of the two peaks $\Delta v_{\mathrm{sep}}$ and the escape fraction of ionising photons $f_{\mathrm{esc}}^{\mathrm{LyC}}$ was estimated using a sample of 56 LyC leaking galaxies at $z<0.4$. Using the peak separation determined in Section \ref{sec:redshift}, and following the relation in \citet{Izotov_18}:
\begin{equation}
    f_\mathrm{esc}^\mathrm{LyC} = \frac{3.23\times10^4}{v_\mathrm{sep}^2} - \frac{1.05\times10^2}{v_\mathrm{sep}} + 0.095\mathrm{~,}
\end{equation}
we find LyC leakage of $f_{\mathrm{esc}}^{\mathrm{LyC}}= 0.18\pm0.08$. The reliability of the double-peak Ly$\alpha$ velocity separation as a proxy of LyC leakage has, however, been called into question. \citet{Naidu_22} have shown, that some sources with high peak separation of $v_\mathrm{sep}>400$ km s$^{-1}$, such as \textit{Ion2, Ion3, Sunburst Arc} and others, leak >$20\%$ of their ionising photons. Additionally, numerical simulations by \citet{Choustikov_23} indicate that while it is unlikely to find strong leakers with wide peak separation, they find no real trend between the two quantities. Similar trend is observed in simulations by e.g. \citet{Giovinazzo_24}, where multiple objects with low peak separation also showcase a low $f_{\mathrm{esc}}^{\mathrm{LyC}}$. However, it is still unclear whether these results are a product of a simulation bias. For this reason, we use additional estimators of LyC leakage to provide reliable measurements. 

\citet{Begley_24} probed the correlation between the escape fraction of Ly$\alpha$ photons $f_{\mathrm{esc}}^{\mathrm{Ly\alpha}}$ and $f_{\mathrm{esc}}^{\mathrm{LyC}}$ with a sample of 152 star-forming galaxies (SFGs) at $4 < z < 5$. They first find a scaling relation between the EW$_0$(Ly$\alpha$) and the Ly$\alpha$ escape fraction, which is then found to be in a linear correlation with $f_{\mathrm{esc}}^{\mathrm{LyC}}$. Following their method, we find that for LAE-11 the escape fraction of Ly$\alpha$ photons is $f_{\mathrm{esc}}^{\mathrm{Ly\alpha}} = 0.53\pm0.06$, and subsequently its LyC leakage is $f_{\mathrm{esc}}^{\mathrm{LyC}}\simeq 0.08\pm0.03$. 

Next, we employ the empirical relation between $f_{\mathrm{esc}}^{\mathrm{LyC}}$ and UV slope $\beta_{1550}$ derived by \citet{Chisholm_22}, using a selection of 89~SFGs and LyC emitters at $z\sim0.3$:
\begin{equation}
    f_{\mathrm{esc}}^{\mathrm{LyC}} = (1.3\pm0.6)\times10^{-4}\times10^{(-1.22\pm0.1)\beta_{1550}}\mathrm{~.}
\end{equation}
Using this relation we calculate $f_{\mathrm{esc}}^{\mathrm{LyC}} = 0.20\pm0.18$. We note that the relation more robustly characterises a population-average escape fractions than an individual scenarios, resulting in the large uncertainty on the calculated $f_{\mathrm{esc}}^{\mathrm{LyC}}$.

Additionally, \citet{Naidu_20} examined the correlation between the star formation rate density $\Sigma_\mathrm{SFR}$ and LyC leakage. We use their empirical model:
\begin{equation}
    f_\mathrm{esc}^\mathrm{LyC} = (1.6\pm0.3)\times\left(\frac{\Sigma_\mathrm{SFR}}{1000~M_\odot\mathrm{yr^{-1}kpc^{-2}}}\right)^{(0.4\pm0.1)}\mathrm{~.}
\end{equation}
We use both $\Sigma_\mathrm{SFR}$ calculated with $L_\mathrm{UV}$ and $L(\mathrm{H}\alpha)$ described in Section \ref{sec:sfr}. Since we only find the lower limits on $\Sigma_\mathrm{SFR}~>~3.59~(9.44)~M_\odot~\mathrm{yr^{-1}~kpc^{-2}}$ for $L_\mathrm{UV}$ ($L(\mathrm{H}\alpha)$), we find the lower limits for $f_{\mathrm{esc}}^{\mathrm{LyC}} > 0.16~(0.24)$ using $L_\mathrm{UV}$ ($L(\mathrm{H}\alpha)$).

To confirm our estimates of $f_{\mathrm{esc}}^{\mathrm{LyC}}\approx (10 - 20) \%$, we use the recent multivariate diagnostic Cox models developed by \citet{Jaskot_24a,Jaskot_24b} to empirically predict $f_\mathrm{esc}^\mathrm{LyC}$ with a combination of observable properties. These models utilise a survival analysis technique to treat data with upper limits appropriately, and was developed using 50 LyC emitting galaxies at $z\sim0.3$. For LAE-11 we use models and their input parameters:
\begin{itemize}
    \item[\textbullet] \textit{LAE} with M$_{1500}$, $E(B-V)_\mathrm{UV}$, and EW$_0$(Ly$\alpha$);
    \item[\textbullet] \textit{ELG-EW} with M$_{1500}$, $\log M_\star$, $E(B-V)_\mathrm{UV}$, and $\log\mathrm{EW_0}$([O{\small III}] + H$\beta$);
    \item[\textbullet] \textit{R50}$-\beta$, which uses M$_{1500}$, $\log M_\star$, $\beta_{1550}$, and the UV half-light radius $R_\mathrm{50,UV}$.
\end{itemize}

\begin{table}[t]
\caption{The parameters and priors used for spectrophotometric SED fitting of LAE-11 with \texttt{BAGPIPES}.}
\centering\label{tab:sed}
    \begin{tabular}{lc}\hline
      Parameter & Prior  \\ \hline\hline
      SFH & non-parametric \citep{Leja_19}\\
      $N_\mathrm{SFH~bins}$ & 7 \\\hline
      dust type & Calzetti \\
      $A_V$ & $(0, 4)$ uniform\\
      $\log~U$ & $(-4, -1)$ uniform\\
      $\log~M_\star/M_\odot$ & $(0, 15)$ uniform\\
      $\log~Z/Z_\odot$ & $(0.0001, 5)$ logarithmic \\
      \hline
    \end{tabular}
\end{table}

Using these models, we find $f_\mathrm{esc}^\mathrm{LyC}~=~(0.33^{+0.36}_{-0.21})$, $(0.18^{+0.38}_{-0.13})$, $(>0.18)$ for models \textit{LAE}, \textit{ELG-EW}, and \textit{R50$-\beta$}, respectively. For \textit{R50$-\beta$} model we only calculate the lower limit of the escape fraction, as the radius $R_\mathrm{UV}$ of the galaxy is smaller than the PSF. We note that the multivariate models point indicate higher uncertainties on the estimated values of the escape fraction. \citet{Jaskot_24b} caution that some galaxies at $z\gtrsim6$ differ from the parameter space covered by the sample of low-$z$ galaxies used to calibrate the model, increasing the uncertainties of the predictions. All the determined $f_{\mathrm{esc}}^{\mathrm{LyC}}$ values and the methods used to estimate them can be found in Table \ref{tab:fesc}. 


We compare the estimated values of $f_{\mathrm{esc}}^{\mathrm{LyC}}$ that were determined using the Ly$\alpha$ emission and those who used other galaxy properties. We find that both the peak separation and \textit{LAE} model indicate a higher escape fraction of LyC ($\gtrsim 15\%$), while the relation by \citet{Begley_24} shows $f_\mathrm{esc}^\mathrm{Ly\alpha}<10\%$. Models and relations focusing on galaxy properties not tied to the Ly$\alpha$ emission line do not show this dichotomy as the $f_{\mathrm{esc}}^{\mathrm{LyC}}-\beta_{1550}$ correlation by \cite{Chisholm_22}, $f_{\mathrm{esc}}^{\mathrm{LyC}}-\Sigma_\mathrm{SFR}$ relations, \textit{ELG-EW} model and the \textit{R50}$-\beta$ model estimate a higher escape fraction of $\gtrsim10\%$. We note, however, that the these relations are either only lower limits on the escape fractions or show a high 1$\sigma$ uncertainty. Comparing these two groups of $f_{\mathrm{esc}}^{\mathrm{LyC}}$ proxies, we find that they do not show a significant scatter among the calculated values, and all indicate that at least $\gtrsim5\%$ of LAE-11's LyC photons are leaking into the surrounding IGM. The agreement between all available indirect tracers indicates that each is viable to determine LyC leakage of high-$z$ galaxies, however multiple tracers are recommended to constrain its exact value. We discuss the implications of the measured $f_{\mathrm{esc}}^{\mathrm{LyC}}$ in Section \ref{sec:burstyEoRgal}.


\begin{figure}[t]
    \centering
    \includegraphics[width=\linewidth]{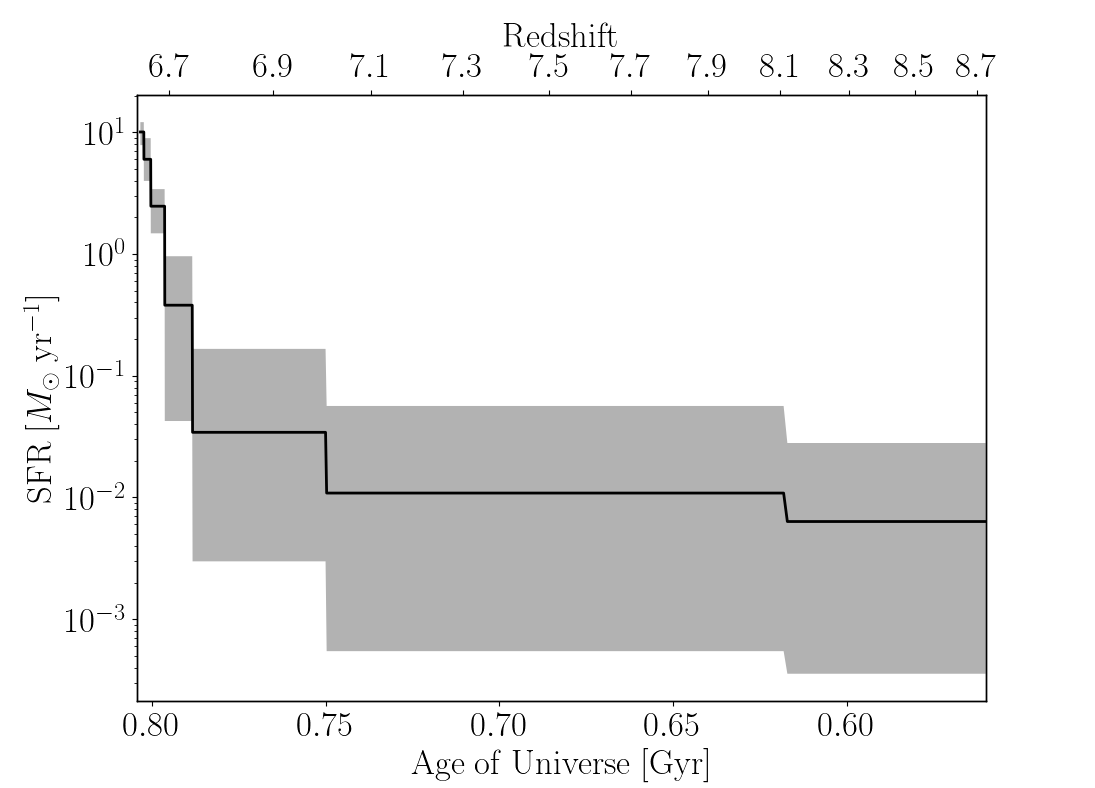}
    \caption{The posterior non-parametric SFH of LAE-11 depicted with solid black line. The 1$\sigma$ uncertainty is marked with a shaded grey region.}
    \label{fig:sfh}
\end{figure}

\subsection{Ionising photon production efficiency $\xi_\mathrm{ion}$}
We calculate the ionising photon production efficiency $\xi_\mathrm{ion}$ of LAE-11 using the H$\alpha$ and H$\beta$ lines, respectively. We have access to both H$\alpha$ and H$\beta$ lines, and therefore we check whether the assumption of Case B recombination scenario is warranted. We find that the ratio of H$\alpha$/H$\beta$ = $2.47\pm0.49$, which is in 1$\sigma$ agreement with Case B scenario of H$\alpha$/H$\beta$ = $2.86$ \citep{Osterbrock_89}. 

In addition to $f_\mathrm{esc}^\mathrm{LyC}$, ionising photon production efficiency needs to be calculated to correctly estimate LAE-11's ionising output. $\xi_\mathrm{ion}$ is defined as:
\begin{equation}\label{eq:xi}
    \xi_\mathrm{ion} =\frac{N(\mathrm{H})}{L_\mathrm{UV,corr}}\frac{1}{1-f_\mathrm{esc}^\mathrm{LyC}} \mathrm{~Hz~erg^{-1},}
\end{equation}
where $N$(H) is the production rate of ionising photons and $L_\mathrm{UV,corr}$ is the intrinsic UV luminosity of the galaxy, calculated from M$_\mathrm{UV}$. We calculate $N$(H) both with H$\alpha$ and H$\beta$ separately, assuming a Case B recombination with the H$\alpha$ (H$\beta$) line coefficient of $C_B = 1.36\times10^{-12}~(4.79\times10^{-13})$ erg, electron density $n_e = 10^3\mathrm{~cm}^{-3}$, and electron temperature of $T_e = 10^4$ K, as $N$(H) = $L_\mathrm{line}/C_B$ \citep{Leitherer_95}. Using this method and $f_\mathrm{esc}^\mathrm{LyC} = 0.18\pm0.08$ estimated from the peak separation of Ly$\alpha$, we measure $\log(\xi_\mathrm{ion}/\mathrm{Hz~erg}^{-1})~= 25.59\pm 0.08$ and $\log(\xi_\mathrm{ion}/\mathrm{Hz~erg}^{-1})~= 25.65 \pm 0.09$ for H$\alpha$ and H$\beta$, respectively. All the measured values and the methods used to calculate them can be found in Table \ref{tab:properties}. We discuss the implications of the measured $\xi_\mathrm{ion}$ in Section~\ref{sec:burstyEoRgal}.

\subsection{Star formation rate and star formation rate density $\Sigma_\mathrm{SFR}$}\label{sec:sfr}
We calculate the star formation rate (SFR) of LAE-11 using its UV and line properties. First, we use the standard relations found in \citet{Kennicutt_98} which uses the \citet{Salpeter_55} IMF and link the SFR of galaxies to their UV luminosity as:
\begin{equation}
    \mathrm{SFR_{UV}} = 1.4\times10^{-28}L_\mathrm{UV}~M_\odot\mathrm{~yr^{-1},}
\end{equation}
and to the luminosity of the H$\alpha$ line following:
\begin{equation}
    \mathrm{SFR_{H\alpha}} = 7.9\times10^{-42}L(\mathrm{H\alpha})~M_\odot\mathrm{~yr^{-1}.}
\end{equation}
We use the UV magnitude and the H$\alpha$ line from the NIRSpec data to calculate the SFR of LAE-11. Following these relations, we find SFR$_\mathrm{UV} = 5.25 \pm 0.69~M_\odot~\mathrm{yr^{-1}}$ and SFR$_\mathrm{H\alpha} = 12.93\pm1.20~M_\odot~\mathrm{yr^{-1}}$. 
All SFR values calculated in this work can be found in Table \ref{tab:properties}. The implications of the measured SFR for LAE-11 are further discussed in Sec.~\ref{sec:sfr_qso}.

\begin{table}[t]
\caption{Estimates of the escape fraction of ionising photons from LAE-11 using indirect empirical relations and multivariate models.}
\centering\label{tab:fesc}
    \begin{tabular}{lcc}\hline
      $f_\mathrm{esc}^\mathrm{LyC}$ & Method & Reference \\ \hline\hline
      $0.18\pm0.08$ & Peak separation & \citet{Izotov_18}  \\
      $0.08\pm0.03$ & $f_\mathrm{esc}^\mathrm{Ly\alpha}$ & \citet{Begley_24}	\\
      $0.20\pm0.18$ & UV slope $\beta$ & \citet{Chisholm_22} \\
      $>0.16$ & $\Sigma_\mathrm{SFR(\mathrm{UV})}$ & \citet{Naidu_20} \\
      $>0.24$ & $\Sigma_\mathrm{SFR(\mathrm{H\alpha})}$ & \citet{Naidu_20} \\
      $0.33^{+0.36}_{-0.21}$ & \textit{LAE} model & \citet{Jaskot_24b}  \\    $0.18^{+0.38}_{-0.13}$ & \textit{ELG-EW} model & \citet{Jaskot_24b} \\ 
      $>0.18$ & \textit{R50}$-\beta$ model & \citet{Jaskot_24b}   \\  \hline
    \end{tabular}
\end{table}

We also evaluate the SFR density of LAE-11 following the definition $\Sigma_\mathrm{SFR} = \mathrm{SFR}/2\pi R_\mathrm{UV}^2$ \citep[e.g.][]{Shibuya_19,Naidu_20}, where $R_\mathrm{UV}$ is the UV effective half-light radius. We measure the FWHM of LAE-11 using the value \texttt{FWHM\_IMAGE}, which was calculated as a part of photometry extraction run with \texttt{SourceXtractor}, where FWHM is calculated assuming a Gaussian core. Next, we pick and stack fifteen bright visually-selected stars. We fit a 2D Gaussian to the stack in order to estimate the FWHM of the PSF, which results in FWHM$_\mathrm{PSF}$ = 0.17\arcsec. However, we find that the FWHM of LAE-11 is lower than this value, and therefore we can only estimate an upper limit on the UV radius. Following the approach of \citet{Torralba_24}, we calculate the UV radius in physical units as FWHM/2, resulting in $R_\mathrm{UV} < 0.47$ kpc. Additionally, we calculate the radius of the galaxy in the Ly$\alpha$ emission present in the \textit{NB926} filter, following the same method. LAE-11 is again unresolved and its FWHM is smaller than the PSF of the narrowband data (FWHM$_\mathrm{PSF}$ = 0.34\arcsec). We can therefore only obtain its upper limits on the Ly$\alpha$ radius of $R_\mathrm{Ly\alpha} < 1.83$ kpc. 

Using the upper limit of $R_\mathrm{UV}$ in turn gives the lower limit for $\Sigma_\mathrm{SFR}$. Using the SFR estimated from UV and from H$\alpha$, we arrive at values $\Sigma_\mathrm{SFR} > 3.59~M_\odot\mathrm{yr^{-1}kpc^{-2}}$ and $> 9.44~M_\odot\mathrm{yr^{-1}kpc^{-2}}$, respectively. Compared to the sample of galaxies at $z\sim6$ \citep{Shibuya_15}, and [O{\small III}] emitters found in the field of COLA1 \citep{Torralba_24}, and strong LyC leakers such as COLA1 \citep{Matthee_18}, \textit{Ion2} and \textit{Ion3} \citep{Vanzella_16, Vanzella_18} and others, LAE-11 belongs in the population with higher $\Sigma_\mathrm{SFR}$ based on the lower limits calculated in this work.

\begin{figure}[t]
    \centering
    \includegraphics[width=\linewidth]{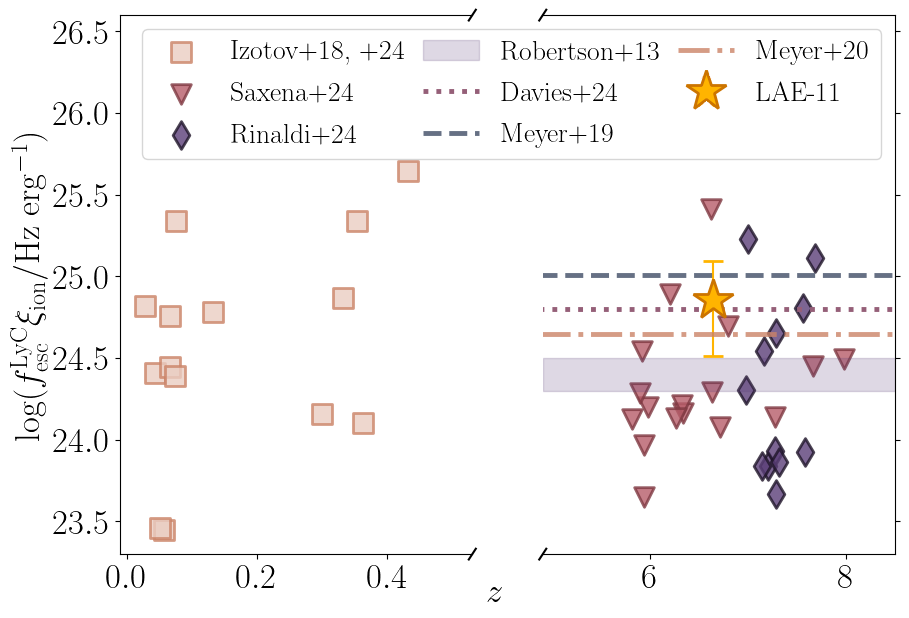}
    \caption{The evolution of the ionising output of galaxies ($\xi_\mathrm{ion}f_\mathrm{esc}^\mathrm{LyC}$) with redshift. The ionising output required to sustain Reionisation determined by \citet{Robertson_13} is depicted with shaded purple region, while the mean ionising output needed to maintain EoR adopted by \citet{Davies_24} is shown with a dotted pink line. The values determined by \citet{Meyer_19,Meyer_20} are shown with a blue dashed and pink dashdotted lines. Low$-z$ SFGs \citep{Izotov_18,Izotov_24} are shown with pale pink squares, high$-z$ faint LAEs \citep{Saxena_24} are shown with red triangles, high$-z$ strong H$\alpha$ emitters \citep{Rinaldi_24} are shown with purple diamonds, and LAE-11 is depicted with a yellow star. We have used the $f_\mathrm{esc}^\mathrm{LyC}$ estimated using the Ly$\alpha$ peak separation.}
    \label{fig:ionoutput}
\end{figure}

\subsection{Inferred properties of quasar J0910-0414}
\begin{figure}[t]
    \centering
    \includegraphics[width=\linewidth]{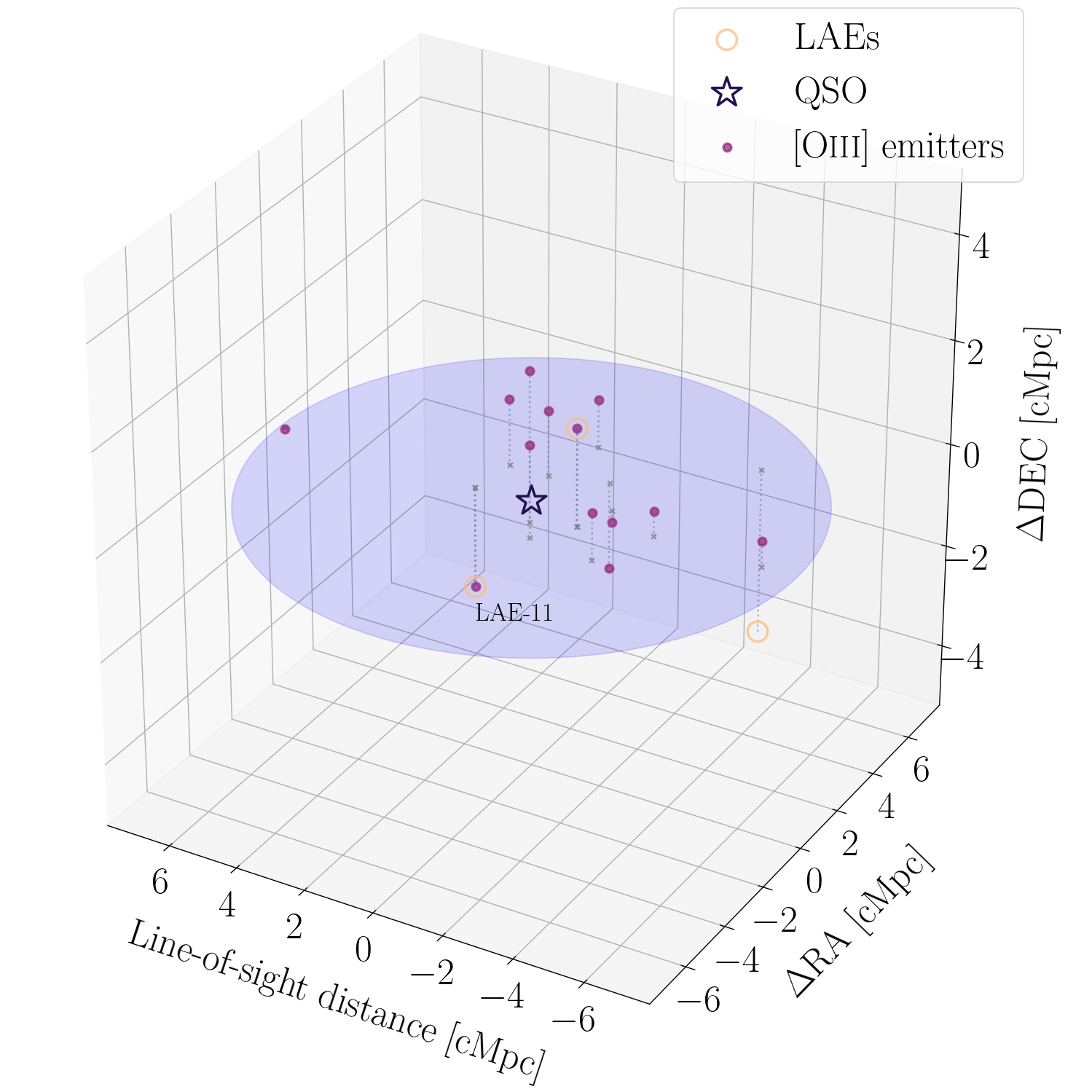}
    \caption{The positions of the LAEs (orange circles) and [O{\tiny III}] emitters (purple dots) in 3D comoving coordinates with respect to the position of J0910-0414 (blue star). The line-of-sight distance is defined to be negative towards the observer with the quasar positioned at 0 cMpc. RA (DEC) is defined to be negative when the RA (DEC) of the galaxy is smaller than the RA (DEC) of the QSO. The purple plane depicts the declination plane of the quasar. Each galaxy has its position projected onto this plane with a gray dashed line.}
    \label{fig:3d}
\end{figure}

Since we are able to observe the blue peak of the Ly$\alpha$ profile at such a high redshift, LAE-11 is certainly residing within the proximity zone of the quasar. Using the LAEs from \citet{Wang_24} and [O{\small III}] emitters found by the ASPIRE survey in the vicinity of the quasar whose positions in 3D space can be seen in Fig. \ref{fig:3d}, we are able to introduce some constraints on the proximity zone of J0910-0414. 

\renewcommand{\arraystretch}{1.65}
\begin{table}[t]
\caption{Physical properties of LAE-11 calculated from line and continuum properties, and SED fitting.}
\centering\label{tab:properties}
    \begin{tabular}{lc}
    \hline
    Property & Value \\ \hline\hline
    $\beta_{1550}$ & $-2.61^{+0.06}_{-0.08}$ \\
    M$_\mathrm{UV,phot}$$^a$ & $-19.84^{+0.14}_{-0.16}$ \\
    M$_\mathrm{UV,cont}$$^b$ & $-19.90^{+0.13}_{-0.14}$ \\ \hline
    $E(B-V)_\mathrm{UV}$ & $0.02_{-0.02}^{+0.03}$ \\ 
    $A_V$ & $0.08_{-0.08}^{+0.14}$ \\
    $A_V$ (\texttt{BAGPIPES})& $<0.01$ \\ \hline
    $\log M_\star/M_\odot$$^c$ & $7.93^{+0.14}_{-0.08}$\\
    $\log U$ &  $-1.33^{+0.21}_{-0.27}$\\
    $\log Z/Z_\odot$ & $-0.86^{+0.03}_{-0.03}$\\
    $R_\mathrm{UV}$ & $<0.47$ kpc \\ 
    $R_\mathrm{Ly\alpha}$ & $<1.83$ kpc \\\hline
    SFR$_\mathrm{UV}$ & $5.25 \pm 0.69~ M_\odot\mathrm{yr}^{-1}$ \\
    SFR$_\mathrm{H\alpha}$ & $12.93\pm1.20~M_\odot\mathrm{yr}^{-1}$ \\
    $\log$ sSFR$_\mathrm{UV}/\mathrm{yr}^{-1}$ & $-7.04\pm0.43$ \\
    $\Sigma_\mathrm{SFR(UV)}$ & $> 3.59 ~M_\odot\mathrm{yr}^{-1}\mathrm{kpc}^{-2}$ \\
    $\Sigma_\mathrm{SFR(\mathrm{H\alpha})}$ & $> 9.44 ~M_\odot\mathrm{yr}^{-1}\mathrm{kpc}^{-2}$\\\hline
    $\log\xi_\mathrm{ion}/\mathrm{Hz~erg}^{-1}$(H$\alpha$) & $25.59\pm0.08$ \\
    $\log\xi_\mathrm{ion}/\mathrm{Hz~erg}^{-1}$(H$\beta$) & $25.65\pm0.09$ \\
    \hline
    \end{tabular}
    \vspace{1ex}
    
     {\raggedright \textbf{Notes.}~$^a$~M$_\mathrm{UV}$ was calculated using the \textit{F115W} flux. $^b$~M$_\mathrm{UV}$ was calculated using the fit to continuum flux.$^c$~Stellar mass $M_\star$ was converted from Kroupa to Salpeter IMF according to \citet{Madau_14}. \par}
\end{table}

\begin{figure}[t]
    \centering
    \includegraphics[width=\linewidth]{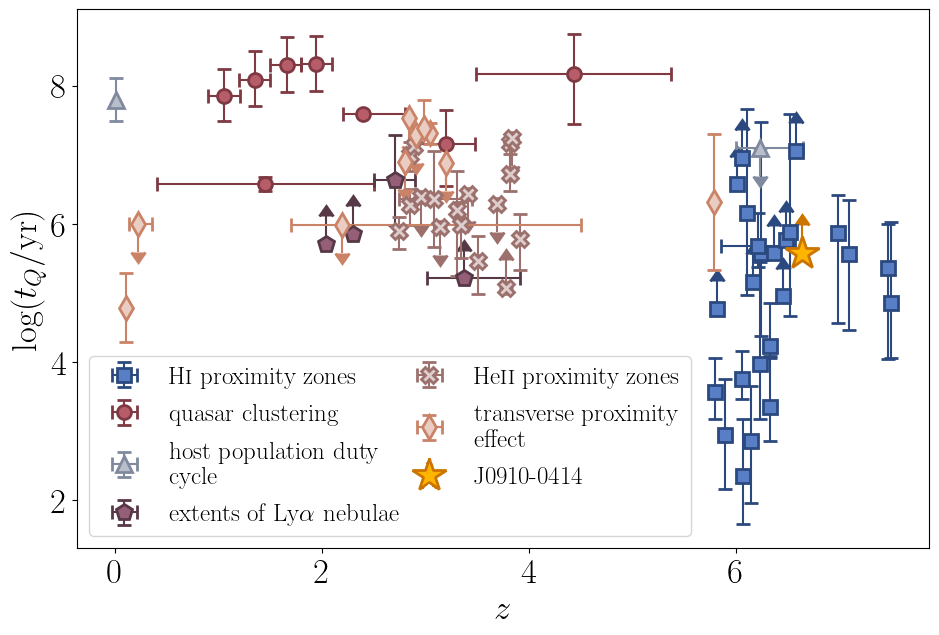}
    \caption{Quasar lifetimes from literature as a function of redshift. Measurements using H{\tiny I} proximity zones along the line of sight are shown with blue squares \citep{Eilers_18, Eilers_21, Davies_19, Davies_20, Andika_20, Morey_21}, studies using quasar clustering are depicted with red circles \citep{Shen_07,Shankar_10,White_12,Laurent_17}. Measurements inferred from the duty cycle of the quasar host population are shown with grey triangles \citep{Yu_02,Chen_18}, while those using extends of Ly$\alpha$ nebulae are depicted with purple pentagons \citep{Trainor_13,Cantalupo_14,Hennawi_15,Borisova_16}. Furthermore, quasar lifetime estimates calculated from the He{\tiny II} proximity zones are shown as light brown crosses \citep{Khrykin_19,Khrykin_21,Worseck_21}, and studies using the transverse proximity effect are marked with a pink diamond \citep{Kirkman_08, Keel_12, Schmidt_17, Schmidt_18, Oppenheimer_18, Bosman_20}. Our estimation of $t_Q$ for J0910-0414 is depicted with a yellow star. Adapted from \citet{Eilers_21}.}
    \label{fig:Qlifetime}
\end{figure}

Quasars are known as intrinsically variable sources. The changes in their brightness and thus production of ionising photons affect the surrounding IGM with a time delay, based on its properties and the distance from the quasar, creating a so-called ``light-echo'' imprint \citep[e.g.][]{Adelberger_04,Visbal_08,Schmidt_19,Kakiichi_22}. By observing the changes in the IGM located further from the quasar, we are able to determine its lifetime $t_Q$. The quasar lifetime is defined such that if the light from the recent activity was observed at time $t = 0$, the quasar actually turned on at a time $-t_Q$ in the past \citep{Eilers_17}. Since LAE-11, located $\sim0.3$ pMpc from the quasar and is residing in the ionised region, it is carrying a imprint of the quasar's past activity. This allows us to determine the lower limit of the quasar's lifetime as defined in \citet{Bosman_20}:
\begin{equation}\label{eq:lifetime}
    t_Q > \frac{\sqrt{d_\parallel^2 + d_\perp^2} - d_\parallel}{c}\mathrm{~s.}
\end{equation}
We measure the line-of-sight proper distance $d_\parallel$ and the perpendicular distance along the plane of the sky $d_\perp$ of LAE-11 from J0910-0414 as $d_\parallel = 0.21\pm 0.01$ pMpc and $d_\perp = 0.25$ pMpc, respectively. Using Eq. \ref{eq:lifetime}, we find the lower limit of J0910-0414's lifetime to be $t_Q > 3.8\times10^5$ yr. 
In Fig.~\ref{fig:Qlifetime}, we compare the lifetime of J0910-0414 to other quasars. 
On the basis of short observed proximity zones along the line-of-sight, some quasars at $z\gtrsim6$ have been identified as being ``young'' with bright lifetimes $t_Q < 10^4$ yr \citep{Eilers_18,Eilers_21}. Even though the proximity zone of J0910-0414 is not visible due to its nature as a BAL quasar, the fact that its ionisation field reaches the location of LAE-11 independently demonstrates that it is not a young quasar. Its bright-phase lifetime is more consistent with the mean lifetime of quasars at $z<5$ measured from their He~{\small{II}} proximity zones \citep{Khrykin_21,Worseck_21} and from clustering arguments (e.g.~\citealt{Laurent_17}).


Next, we analyse J0910-0414's possible opening angles. We assume that the quasar is radiating in a bipolar cone with an opening angle $\theta_\mathrm{Q}$ with an ionising photon production rate $\dot{N}_\mathrm{ion}^\mathrm{QSO}$. By using the available photometry in the \textit{y}$_{ps1}$ band and the UV slope~$\beta$ for J0910-0414 from \citet{Yang_21}, we integrate over the luminosity density $L_\nu$ of the quasar as $\dot{N}_\mathrm{ion}^\mathrm{QSO}~=~\int^{\nu_\mathrm{HeII}}_{\nu_\mathrm{HII}}~L_\nu/h\nu ~\mathrm{d}\nu$. We find that the ionising photon production rate in the observed frame of the quasar is $\dot{N}_\mathrm{ion}^\mathrm{QSO}~=~(3.43\pm0.02)\times10^{57}$~s$^{-1}$. The quasar's surroundings are unaffected by the photoionisation rate outside of the bipolar cones, while on the inside they are affected by a ionisation field given by photoionisation rate $\Gamma_\mathrm{HI}^\mathrm{QSO}$:
\begin{equation}
    \Gamma_\mathrm{HI}^\mathrm{QSO}(d_\perp, d_\parallel) = \frac{-\sigma_{\lambda912}\beta}{3-\beta}\frac{\dot{N}_\mathrm{ion}^\mathrm{QSO}[-\Delta t(d_\perp, d_\parallel)]}{4\pi(d_\perp^2 + d_\parallel^2)}\mathrm{~s^{-1}~,}
\end{equation}
where $\sigma_{\lambda912}$ is the photoionisation cross section at the Lyman Limit ($\lambda = 912$ \AA), $\Delta t(d_\perp, d_\parallel)$ is the time delay at a distance $d_\mathrm{QSO}^2 = d_\perp^2 + d_\parallel^2$ from the quasar, $d_\perp$ is the perpendicular separation along the plane of the sky and $d_\parallel^2$ is the line-of-sight distance from the quasar with a negative sign towards the observer. Due to the tight constraints on the redshifts of both the quasar and LAE-11, we locate the galaxy behind the quasar at $230.93\degrees$$^{+1.31}_{-1.36}$ away from the observer in counter-clockwise direction. This position allows us to constrain the minimal opening angle of the quasar $\theta_\mathrm{Q}$, since both the observer and LAE-11 are within the cone. More specifically, the BAL nature of J0910-0414 suggests that we are viewing it at an angle very close to the edge of the ionisation cone (e.g.~\citealt{DiPompeo_12,Nair_22}). 

\begin{figure}[t]
    \centering
    \includegraphics[width=\linewidth]{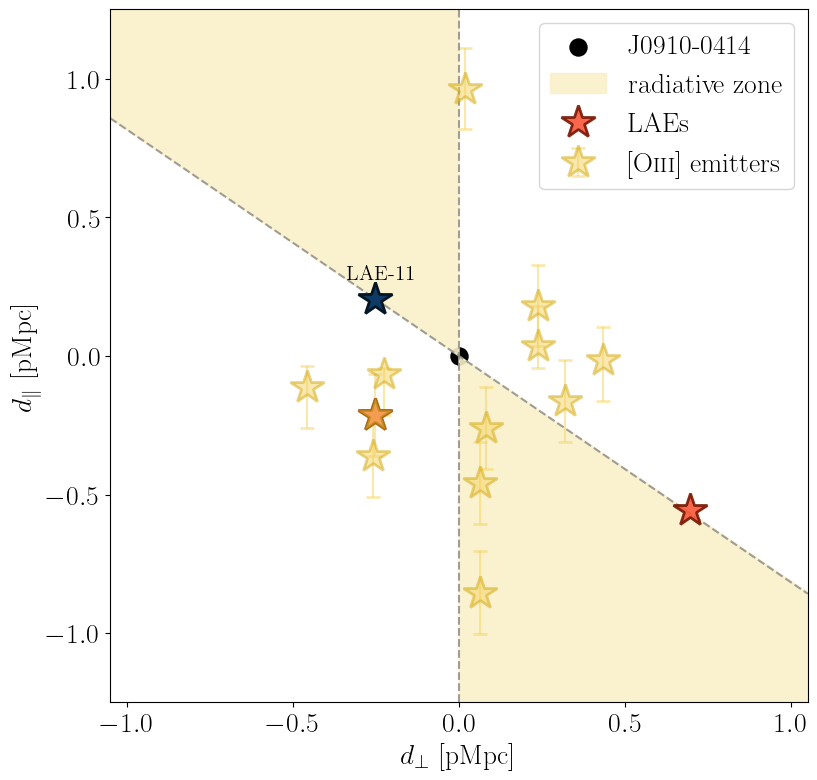}
    \caption{The physical neighbourhood of Q9010 (black circle) surrounded by proximate [O{\tiny III}] emitters from the ASPIRE survey (yellow stars) and LAEs from \citet[][red stars]{Wang_24}, and LAE-11 (dark blue star). The parallel distance $d_\parallel$ is defined as negative towards the observer. The perpendicular distance $d_\perp$ is defined negative if $\mathrm{RA}_\mathrm{gal} < \mathrm{RA}_\mathrm{QSO}$. We note that LAE-12 is also an [O{\tiny III}] emitter, and showed with an orange colour. J0910-0414's ionisation field (yellow shaded region) is depicted ``counter-clockwise'' from the observer and depicts the lower limit on its opening angle $\theta_Q$, assuming a bipolar cone geometry. Since both the observer and the double-peaked LAE must be within this region, we can infer that the lower limit of $\theta_Q > 49.62\degrees$.}
    \label{fig:qso_field}
\end{figure}

We note that the other two LAEs observed in close proximity to J0910-0414 (red and dark orange in Fig.~\ref{fig:qso_field}) do not show double-peaked Ly$\alpha$ morphologies, but this is not necessarily an indication that they are located \textit{outside} of the ionisation cone; at $z<3$, only roughly $\sim1/3 - 1/2$ of LAE are double-peaked \citep{Kulas_12, Trainor15}. LAE-11, therefore, needs to be situated in either of the two cones (``clockwise'' or ``counter-clockwise'' from the observer). Using the 1$\sigma$ value of LAE-11's positional angle, the opening angle of the quasar $\theta_\mathrm{Q}$ is then constrained to
\begin{align}
\theta_\mathrm{Q} > \arctan\left(\frac{r_{\perp,~\mathrm{LAE-11}}}{r_{\parallel,~\mathrm{LAE-11}}}\right) > 49.62 \degrees &&\lor&& \theta_\mathrm{Q} > 130.38\degrees\mathrm{~,}
\end{align}
for ``counter-clockwise'' and ``clockwise'' direction as shown in Fig.~\ref{fig:qso_field}. The only other quasar for which this measurement could be performed, J0836+0054 \citep{Bosman_20}, requires an ionisation opening angle $\theta_\mathrm{Q} > 21\degrees$ to illuminate its proximate double-peaked LAE (see also \citealt{Borisova_16}). For both of these quasars, the ionisation opening angle is therefore much larger than the measured width of quasar radio jets, $\theta_\mathrm{Q, radio} < 2\degrees$  \citep{Pushkarev17}.

\section{Discussion}\label{sec:discussion}
\subsection{Quasar contribution to LAE-11's Ly$\alpha$ emission}
The ionisation field of QSOs can influence the Ly$\alpha$ emission of its nearby LAEs, contributing with a \textit{fluorescence-triggered} emission. Fluorescence results from the reprocessing of ionised photons from the QSO by the neutral hydrogen gas clouds in the proximate galaxies \citep[e.g.][]{Hogan_87, Cantalupo_07, Hennawi_13}. This effect is observable either if the LAE is in the foreground, or $d \lesssim 3.2$ pMpc in the background of the quasar \citep{Trainor_13}. Since LAE-11 is only $\sim0.3$ pMpc behind J0910-0414, we calculate the quasar's contribution to its Ly$\alpha$ luminosity. 

Following the method of \citet{Bosman_20}, we define the fluorescent Ly$\alpha$ luminosity contributed to a system with a 
cross section $\sigma_\mathrm{LAE}$ and distance $d_\mathrm{QSO}$ from the quasar as:
\begin{equation}
    L_\mathrm{fluor.}(\mathrm{Ly\alpha}) \simeq \frac{2}{3}h\nu_\alpha f_\phi\frac{\sigma_\mathrm{LAE}}{4\pi d_\mathrm{QSO}^2}\dot{N}_\mathrm{ion}^\mathrm{QSO}\mathrm{~erg~s^{-1},}
\end{equation}
where $f_\phi$ is the illuminating fraction. Assuming that all fluorescent emission is scattered into the line-of-sight, the illuminating fraction takes on the value $f_\phi = 1$. For the cross section of LAE-11, we assume a simplified spherical geometry of $\sigma_\mathrm{LAE-11} = 4\pi R^2_\mathrm{Ly\alpha}$. Finally, we use $\dot{N}_\mathrm{ion}^\mathrm{QSO}$ calculated in the previous section and find that J0910-0414 contributes $<3\%$ within 3$\sigma$ uncertainty to the Ly$\alpha$ luminosity of LAE-11. 

\subsection{Star formation in the quasar environment}\label{sec:sfr_qso}
Quasars and active galactic nuclei (AGNs) have been proposed as a likely driver behind quenching of star formation in host and proximate galaxies, as their outflows can prevent gas from cooling and delay the onset of star formation \citep[e.g.][]{Silk_98, Kashikawa_07, Fabian_12} or even completely expel gas reservoirs in small haloes ($M_h \leq 10^7~M_\odot$) located at a distance of 1 pMpc in $\lesssim10^8$ years \citep{Shapiro_04}. Due to LAE-11's close proximity to J0910-0414, we wish to examine the impact of the quasar's ionisation field on the star formation of the galaxy.   

We begin by calculating the UV intensity of the QSO radiation at the location of LAE-11 in the units of $J_{21}$, where $J_{21} = 1$ corresponds to an intensity of $10^{-21}~\mathrm{erg~s^{-1}cm^{-2}Hz^{-1}sr^{-1}}$ at the Lyman limit (LL; $\lambda = 912$ \AA):
\begin{equation}
    \frac{L_\mathrm{LL}^\mathrm{QSO}}{(4\pi d)^2} = J_{21}(d)\times10^{-21}~\mathrm{erg~s^{-1}cm^{-2}Hz^{-1}sr^{-1}~,} 
\end{equation}
where $L_\mathrm{LL}^\mathrm{QSO}$ is the quasar's luminosity at the Lyman limit and $d$ is the physical distance from the quasar to LAE-11. We again utilise the available photometry in the $y_{ps1}$ band and the UV slope $\beta$ from \citet{Yang_21} to calculate $L_\mathrm{LL}^\mathrm{QSO}$. Following this method, we find that the UV intensity at LAE-11's location is $J_{21} = 18.63\pm2.04$. \citet{Kashikawa_07} show that star formation is completely suppressed in haloes with $M_\mathrm{h} \leq 3\times10^9~M_\odot$ at intensity $J_{21} \geq 3$. At intensities of $J_{21}\sim20$, star formation is delayed by $\sim50$ Myr in halos with $M_\mathrm{h} \leq 10^{10}~M_\odot$. The absence of a Balmer break in the spectrum of LAE-11 points towards an absence of older stellar population. This trend can also be observed in the recovered SFH from the spectrophotometric SED fit (see Fig. \ref{fig:sfh}). We note, however, that \citet{Witten_24} found that a young stellar population of 1/3 of the total stellar mass can effectively suppress the Balmer break in the observed spectra of high-$z$ galaxies. The fact that LAE-11 appears to have formed recently is consistent with a delay in the star formation due to the quasar's influence, but it is also possible that the galaxy's recent growth would have occurred even in the quasar's absence. Indeed, studies of larger samples of galaxies around $z>6$ quasars find that the quasar has a negligible impact on their properties \citep{Champagne_24a,Champagne_24b}. Therefore, we can only infer the lower limit on the halo mass following the \citet{Kashikawa_07} model as $3\times10^9~M_\odot \lesssim M_h$.



\subsection{LAE-11 as a bursty Reionisation-driving galaxy}\label{sec:burstyEoRgal}
LAE-11 is a galaxy with a blue UV slope $\beta_{1550}$ ($\beta_{1550}=-2.61^{+0.06}_{-0.08}$). \citet{Topping_24} have investigated the UV slope of galaxies at high redshift and found the median value of $\beta\sim -2.3$ for $5 < z < 7.3$. However, they also find 44 objects with extremely blue UV slopes ($\beta\leq -2.8$), which are best described by density bounds H{\small II} regions with $f_\mathrm{esc}^\mathrm{LyC}\sim0.5$. LAE-11 is comparable to these outliers rather than the average galaxy at these redshifts, indicating a high LyC escape fraction. This is further confirmed by our calculation using the $f_\mathrm{esc}^\mathrm{LyC}-\beta$ relation found by \citet{Chisholm_22}. The significant escape fraction of LyC photons is further confirmed with a selection of indirect tracers (see Table \ref{tab:fesc}), which all point towards an $f_\mathrm{esc}^\mathrm{LyC}$ of at least $>5\%$. 

The effective production of ionising photons of LAE-11 (Table \ref{tab:properties}) is comparable to other blue star-forming galaxies at high-$z$ \citep[e.g.][]{Bouwens_16, Nakajima_18, Saxena_24, Simmonds_23, Simmonds_24, Torralba_24} as well as H$\alpha$ emitters (HAEs) at $z\sim 7 - 8$ \citep{Rinaldi_24}. Additionally, this value is in line with the ionising photon production efficiency of low-$z$ LyC leakers \citep[e.g.][]{Izotov_18,Izotov_24}. However, $\log(\xi_\mathrm{ion}/\mathrm{Hz~erg^{-1}})$ of LAE-11 is higher than the average value found for faint galaxies at $z < 4$ \citep{Bouwens_16}, or HAEs at $z\sim 2 - 3$ \citep{Chen_24}.

By measuring both the $f_\mathrm{esc}^\mathrm{LyC}$ and $\xi_\mathrm{ion}$ of LAE-11, we can investigate its total ionising output, which results in $\log(f_\mathrm{esc}^\mathrm{LyC}\xi_\mathrm{ion,H\alpha}/\mathrm{Hz~erg^{-1}})~=~24.85^{+0.24}_{-0.34}$. Compared to low-$z$ SFGs, high-$z$ faint LAEs and high-$z$ H$\alpha$ emitters, LAE-11 is compatible with the galaxies with effective ionising output (see Fig. \ref{fig:ionoutput}). Additionally, the usually assumed ionising output of galaxies required to maintain the reionisation of the Universe is $\log(f_\mathrm{esc}^\mathrm{LyC}\xi_\mathrm{ion}/\mathrm{Hz~erg^{-1}}) = 24.3 - 24.6$ \citep{Robertson_13}, $25.01$ \citep{Meyer_19}, $24.64$ \citep{Meyer_20} or $24.8$ \citep{Davies_24}, which are lower or comparable to LAE-11's value. LAE-11 could therefore be a prototypical galaxy with a significant contribution to the ionising budget near the end of the EoR.

The blue UV slope, and an absence of a Balmer break possibly point towards a very young stellar population of the galaxy. The SFR of LAE-11 estimated using the UV emission is lower compared to the one diagnosed via the H$\alpha$ line with SFR$_\mathrm{H\alpha}$/SFR$_\mathrm{UV} = 2.46\pm0.40$. While the nebular emission of H$\alpha$ traces the formation of the most massive O and B stars with lifetimes of a few million years, the UV emission arises from a wider mass range of the same stellar population measuring timescales of 100 Myr \citep[e.g.][]{Lee_09}. Therefore, the difference in these two tracers might indicate that the galaxy is undergoing a burst of star formation at the time of the observation \citep[e.g.][]{Glazebrook_99, Sullivan_00, Dominguez_15, Emami_19, Asada_24}. 
This is further supported by the SFH extracted by SED fitting (Fig. \ref{fig:sfh}).

Reionisation-era galaxies are characterised by extreme [O{\small III}] and H$\beta$ emission, with equivalent widths ranging from 300~\AA~to 3000~\AA~\citep[e.g.][]{Endsley_21, Endsley_23}, as expected in galaxies with young stellar populations and low metallicity \citep[e.g.][]{Labbe_13,DeBarros_19}. LAE-11 meets all of these conditions, with $\mathrm{EW_0}$([O{\small III}] + H$\beta$) $=1843.94\pm192.83$~\AA~and $\log Z/Z_\odot=-0.86\pm0.03$. Furthermore, the high equivalent widths of Ly$\alpha$, [O{\small III}] and H$\beta$ in the spectra of LAE-11 are comparable to the most extreme EELGs at $z\sim2-3$ identified by \citet{Tang_24} (see Fig. \ref{fig:o3hbvpeak}). 

\subsection{Alternative interpretations of LAE-11 at $z_\mathrm{[OIII]}$}\label{sec:o3redshift}
In this subsection we interpret the ionising properties of LAE-11 when the $z_\mathrm{[OIII]}$ is taken at face-values. At this redshift, the expected Ly$\alpha$ emission line falls on the left peak of LAE-11's Ly$\alpha$ profile. The velocity offset the left peak from the systemic redshift in this scenario falls to $v_\mathrm{left} = 4.80\pm24.27$ km s$^{-1}$ and the velocity offset of the right peak is $v_\mathrm{right} = 260.75\pm24.44$ km s$^{-1}$. A small sample of galaxies with a similar profile (red-dominated with blue peak situated at systemic redshift) were found at lower redshift by \citet{Kulas_12}, \citet{Trainor15} and \citet{Vitte_24}. These works suggest that the origin of the double peak in this case arises from processes other than radiative transfer of Ly$\alpha$ photons. Since LAE-11 is a compact source and neither of the peaks is spatially offset from the other in the 2D spectra, we discard the possibility of a satellite source or merger contaminating our data. 

Barring this scenario, the most likely interpretation of the profile would then position the apparent left peak into the centre position of a triple-peaked intrinsic line profile. In this case, the central peak corresponds to Ly$\alpha$ emission which escapes without scattering through holes in the neutral interstellar medium (ISM), while the other two peaks arise due to Ly$\alpha$ photon scattering. In order to see this peak, the IGM around LAE-11 must still be ionised to a high degree. This interpretation is supported by the weak detection of transmission in the spectrum at $v_\text{sep}\sim-400$ km s$^{-1}$ at $\lesssim2\sigma$, which could be interpreted as the third Ly$\alpha$ line peak (see Fig.~\ref{fig:LyaProf}). 
Since the left peak would act as the central peak of the line, and is observed, there is an ionised channel in the ISM allowing a 100\% leakage of LyC photons from the galaxy \citep[e.g.][]{Zackrisson_13}. However, this result of $f_\mathrm{esc}^\mathrm{LyC} = 100\%$, would indicate that the ionising photon production efficiency $\xi_\mathrm{ion}$ approaches infinity as indicated by Eq.~\ref{eq:xi}, since no Balmer emission lines are expected in this scenario. We also remind the reader of the discussion in Section~\ref{sec:redshift}, where discrepancies are identified between the Ly$\alpha$ and [O~{\small{III}}] redshifts not only for LAE-11 but also for another galaxy which similarly falls on module A of the NIRCam observations.

A final alternative interpretation of the line shape, if the $z_\mathrm{[OIII]}$ is taken at face value, could be the presence of a neutral hydrogen gas clump in the vicinity of the galaxy. This clump would be responsible for the trough feature in the line profile, which in this scenario has a velocity offset of $v_\mathrm{trough} = 125.25\pm22.06$ km s$^{-1}$. This gas with neutral hydrogen column density of $\log N_\mathrm{HI} = 13.96\pm0.24$ would therefore be falling into the galaxy. Since the double-peak in this scenario would not result from Ly$\alpha$ radiative transfer, the peak separation could not be used for LyC leakage calculation and only other proxies would have to be utilised. However, even in this scenario, LAE-11 would still be required to reside in an ionised IGM as the Ly$\alpha$ emission line would still be highly transmissive at the systemic velocity.

\section{Summary}
We present an analysis of a double-peaked Ly$\alpha$ emitter in the proximity of quasar J0910-0414 at $z = 6.6405\pm0.0005$ discovered in a search for a protocluster anchored by a high-$z$ quasar by \citet{Wang_24}. The detection of the blue peak implies that the galaxy is located within the ionising radiation of the quasar's proximity zone. We utilise a combination of ground based high-resolution Keck/DEIMOS spectroscopy, JWST WFSS spectroscopy, low-resolution NIRSpec MSA spectroscopy, as well as photometric data spanning the UV to optical wavelengths to characterise the physical parameters and ionising output of the galaxy. Moreover, we use our findings to introduce constraints on the properties of the central quasar. Our key findings are summarised as follows:
\begin{enumerate}
    \item LAE-11 displays a narrow double-peaked Ly$\alpha$ emission profile with a velocity separation of $\Delta v_\mathrm{sep} = 255.71\pm33.69$~km/s. The peak separation is comparable to luminous low redshift LAEs and GPs, while high-$z$ double-peaked LAEs \citep{Hu_16,Songaila_18,Bosman_20,Meyer_21} tend to represent the brightest sources with wider peak separation.
    \item LAE-11 is a fairly bright UV galaxy (M$_\mathrm{UV} = -19.84^{+0.14}_{-0.16}$) with a moderately steep UV slope ($\beta_{1550} = -2.61^{+0.06}_{-0.08}$). Analysing the direct imaging of LAE-11 reveals a very compact structure of $R_\mathrm{UV}<0.47$ kpc. Using the Balmer decrement, indirect tracers, and SED fitting, we find that LAE-11 is a very dust-poor system. The spectrophotometric fit furthermore characterises the galaxy as low mass ($\log(M_\star/M_\odot) = 7.93^{+0.14}_{-0.08}$) with low metallicity ($\log(Z/Z_\odot) = -0.86\pm0.03$). We are also able to report on the SFR of the galaxy determined both by UV continuum and H$\alpha$ line to be SFR$_\mathrm{UV} = 5.25\pm0.69~M_\odot\mathrm{~yr^{-1}}$ and SFR$_\mathrm{H\alpha} = 12.93\pm1.20~M_\odot\mathrm{~yr^{-1}}$. 
    \item Utilising a number of indirect tracers and multivariate models, we evaluate the escape fraction of ionising photons $f_{\mathrm{esc}}^{\mathrm{LyC}}$. We find that all tracers point towards an escape fraction of at least $\geq5\%$, with all diagnostics pointing towards an escape fraction of $f_{\mathrm{esc}}^{\mathrm{LyC}} = (0.08-0.33)$. Furthermore, we calculate the ionising photon production efficiency to be $\log(\xi_\mathrm{ion}/\mathrm{Hz~erg^{-1}}) = 25.59\pm0.08$ ($25.63\pm0.08$) using the H$\alpha$ (H$\beta$) emission line detected in the NIRSpec spectrum. The total ionising output of the galaxy is therefore $\log(f_\mathrm{esc}^\mathrm{LyC}\xi_\mathrm{ion,H\alpha}/\mathrm{Hz~erg^{-1}}) = 24.85^{+0.24}_{-0.34}$. This is higher or comparable to the canonical value commonly used for galaxies in order to maintain the EoR \citep{Robertson_13,Meyer_19,Meyer_20,Davies_24}. LAE-11 therefore belongs to the population of galaxies which inject a significant amount of ionising radiation into the IGM in the early Universe.
    \item The star formation in LAE-11 has not been extinguished despite the quasar induced ionising intensity of $J_{21} = 18.63\pm2.04$, conversely it is undergoing an active star burst at the time of the observation. Therefore, the burst in SFR and the possible lack of older stellar population points towards a recent formation of the galaxy. This is consistent with a delayed SFR due to QSO influence, however studies of larger samples of of galaxies around $z>6$ quasars indicate that their impact on surrounding galaxies is insignificant \citep{Champagne_24a,Champagne_24b}. Following the model presented in \citet{Kashikawa_07} implies that mass of its dark matter halo is $3\times10^9 M_\odot \lesssim M_h$.
    \item Finally, we constrain the opening angle and lifetime of the central quasar J0910-0414. By constraining the location of LAE-11, we found the lower limit on the quasar's opening angle to be $\theta_Q>49.62\degrees$ and the lower limit on the quasar lifetime to be $t_Q >3.8\times10^5$ years.
\end{enumerate}


\begin{acknowledgements}
      KP and SEIB are supported by the Deutsche Forschungsgemeinschaft (DFG) under Emmy Noether grant number BO 5771/1-1. 
      Data presented herein were obtained at Keck Observatory, which is a private 501(c)3 non-profit organization operated as a scientific partnership among the California Institute of Technology, the University of California, and the National Aeronautics and Space Administration. The Observatory was made possible by the generous financial support of the W. M. Keck Foundation. This research is furthermore based in part on data collected at the Subaru Telescope, which is operated by the National Astronomical Observatory of Japan. The authors wish to recognize and acknowledge the very significant cultural role and reverence that the summit of Maunakea has always had within the Native Hawaiian community. We are most fortunate to have the opportunity to conduct observations from this mountain.
      Moreover, this research is based on observations made with the NASA/ESA Hubble Space Telescope obtained from the Space Telescope Science Institute, which is operated by the Association of Universities for Research in Astronomy, Inc., under NASA contract NAS 5–26555. These observations are associated with program \#16187. This work is also based in part on observations made with the NASA/ESA/CSA James Webb Space Telescope. The data were obtained from the Mikulski Archive for Space Telescopes at the Space Telescope Science Institute, which is operated by the Association of Universities for Research in Astronomy, Inc., under NASA contract NAS 5-03127 for JWST. These observations are associated with program \#2078.
\end{acknowledgements}

%
\bibliographystyle{aa} 
\bibliography{bibliography} 

\begin{appendix} 
\onecolumn
\section{Additional figures}
\begin{figure*}[h]
\centering
\includegraphics[width=0.8\textwidth]{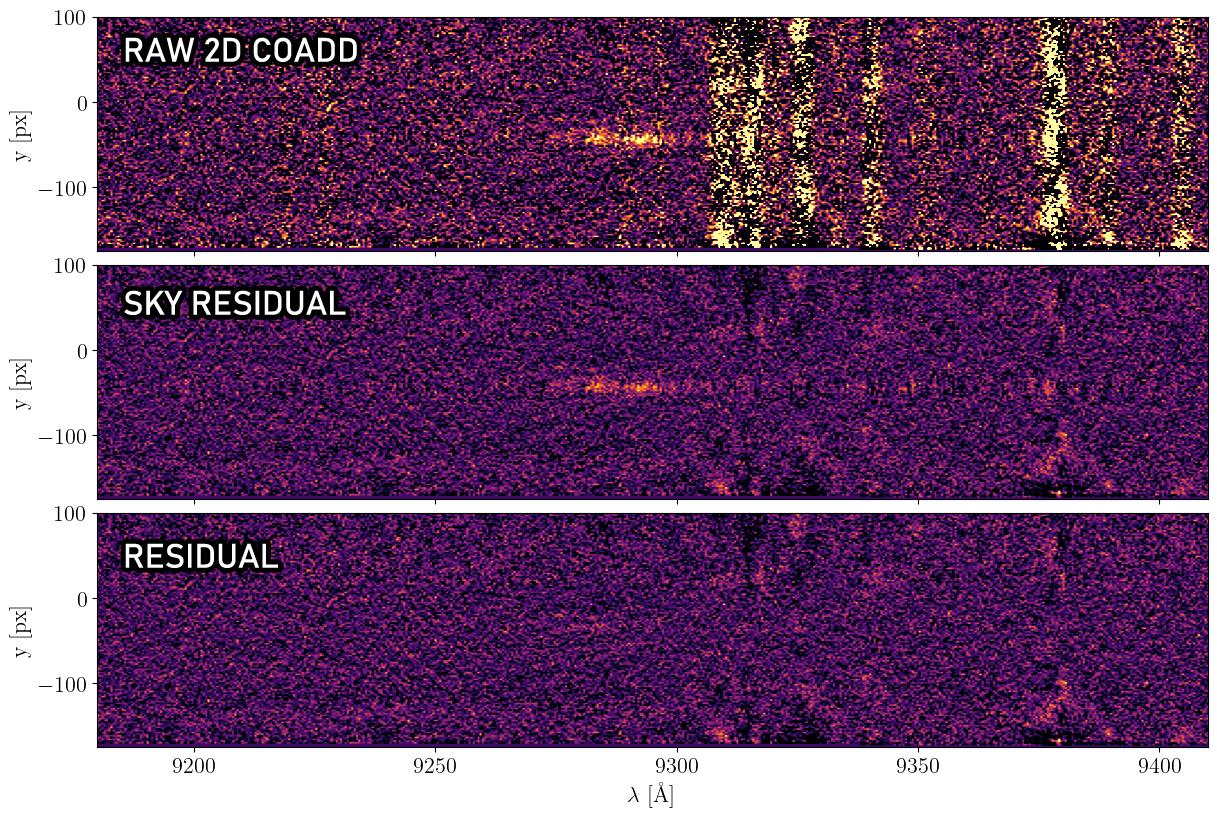}
\caption{\textit{(upper)} The raw coadded 2D spectrum generated by the \texttt{Pypeit} pipeline. \textit{(middle)} The sky subtraction residuals resulting from the subtraction of the sky model from the raw coadded 2D spectrum. \textit{(lower)} The final residuals after subtracting both the sky model and the object model, which is used to extract the 1D science spectrum.}
\label{fig:skysub}
\end{figure*}

\begin{figure*}[h]
\centering
\includegraphics[width=0.8\textwidth]{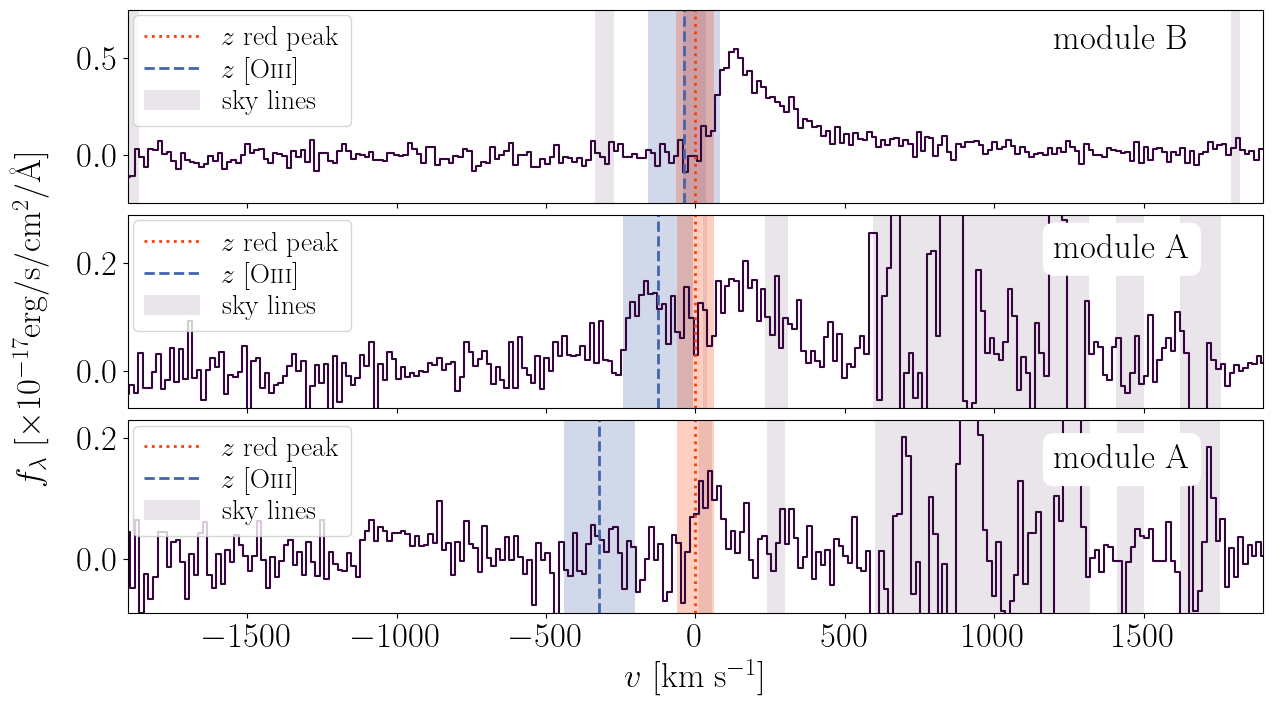}
\caption{DEIMOS spectra of the LAEs in our sample with [O{\tiny III}] spectra from \textit{JWST}/NIRCam WFSS. LAE-1 \textit{(upper)}, LAE-11 \textit{(middle)} and LAE-12 \textit{(lower)} are shown with a purple line, and the skyline regions are marked with a shaded purple region. The dashed blue line represents the redshift of the galaxy as measured from the [O{\tiny III}] doublet with the shaded blue region representing the conservative calibration redshift offset adopted by \citet{Wang_23}, while the dotted orange line corresponds to the redshift calculated using FWHM(Ly$\alpha$) as seen in Eq. \ref{eq:fwhm} with the shaded orange region representing its measured uncertainty.}
\label{fig:module}
\end{figure*}

\begin{figure*}
\includegraphics[width=\textwidth]{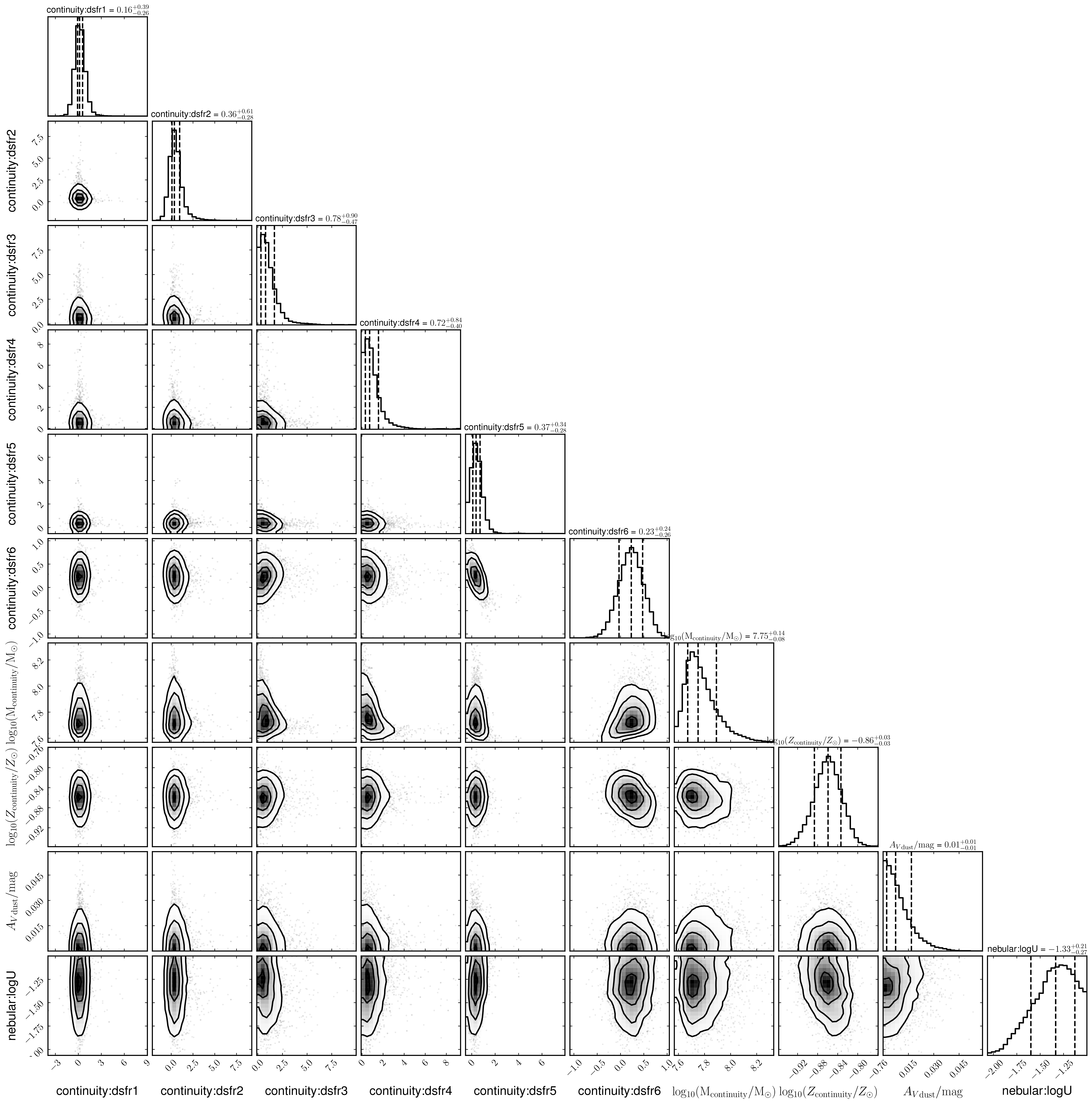}
\caption{Corner plots showing the posterior distributions of the free parameters of the SED fit of LAE-11 using \texttt{BAGPIPES}. The contours indicate the 1$\sigma$, 2$\sigma$, and 3$\sigma$ posterior of each parameter. The histogram of each parameter is shown on top of the corner plots, with the median shown with a dashed lines indicating the median and 1$\sigma$ distribution.}
\label{fig:bagpipesPosterior}
\end{figure*}
\end{appendix}

\end{document}